\documentclass[a4paper,11pt]{article}
\pdfoutput=1 

\usepackage{jcappub} 

\usepackage[T1]{fontenc} 

\usepackage[utf8]{inputenc}
\usepackage{enumerate}
\usepackage{graphicx}
\usepackage{booktabs}
\usepackage{tabularx}
\usepackage{textcomp}
\usepackage{xspace}
\usepackage{amsmath}
\usepackage{amsfonts}
\usepackage{amssymb}
\usepackage[dvipsnames,table]{xcolor}

\hypersetup{urlcolor=BlueViolet,
	    citecolor=Plum,
	    linkcolor=PineGreen}
\usepackage[official]{eurosym}
\definecolor{lightgray}{rgb}{0.9,0.9,0.9}	    
\definecolor{green}{rgb}{0,0.5,0}
\definecolor{red}{rgb}{1,0,0}
\definecolor{blue}{rgb}{0,0,0.5}

\newcommand{\exclude}[1]{}

\title{Miniclusters from axion string simulations}

\author[a,b]{Giovanni Pierobon,}
\author[c,d]{Javier Redondo,}
\author[e]{Ken'ichi Saikawa,}
\author[c]{Alejandro Vaquero}
\author[f]{and Guy D.\ Moore}

\affiliation[a]{School of Physics, University of New South Wales, Sydney NSW 2052, Australia}
\affiliation[b]{Sydney Consortium for Particle Physics and Cosmology, Sydney NSW 2052, Australia}
\affiliation[c]{CAPA \& Departamento de Fisica Teorica, Universidad de Zaragoza, 50009 Zaragoza, Spain}
\affiliation[d]{Max-Planck-Institut f\"ur Physik (Werner-Heisenberg-Institut), F\"ohringer Ring 6, 80805 M\"unchen, Germany}
\affiliation[e]{Institute for Theoretical Physics, 
			Kanazawa University,
			Kakuma-machi, Kanazawa, 
			Ishikawa 920-1192, 
			Japan}
\affiliation[f]{Institut f\"ur Kernphysik, Technische Universit\"at Darmstadt,
Schlossgartenstra{\ss}e 2, D-64289 Darmstadt, Germany}

\emailAdd{g.pierobon@unsw.edu.au}
\emailAdd{jredondo@unizar.es}
\emailAdd{saikawa@hep.s.kanazawa-u.ac.jp}
\emailAdd{alexv@unizar.es}
\emailAdd{guy.moore@physik.tu-darmstadt.de}

\abstract{The properties of axion miniclusters and of the voids between them can have very strong implications for the discovery of axions and the dark matter of the Universe. These properties can be strongly affected by axion dynamics in the early Universe, such as the axion string network and the non-linear dynamics around the QCD phase transition. Recently, improvements in numerical simulation techniques have allowed us to calculate the dark matter axion field from axion strings and QCD effects using different methods: directly with low-tension strings but high resolution~\cite{Vaquero:2018tib},  directly with effective high-tension strings~\cite{Klaer:2017ond}, or indirectly by extrapolating an attractor solution~\cite{Gorghetto:2020qws}. In this work, we study the properties of miniclusters in the different approaches used in the literature. We find that, while there are substantial differences in the mass distribution and internal density profiles, globally there is a similar energy distribution between minicluster halos and voids.}

\begin{document}

\maketitle
\flushbottom

\section{Introduction}

Originally introduced as a solution to the strong-CP puzzle of quantum chromodynamics (QCD) \cite{Peccei:1977hh}, the \emph{axion} is one of the most compelling dark matter (DM) candidates \cite{Bertone:2016nfn,Preskill:1982cy,Abbott:1982af,Dine:1981rt,Dine:1982ah,Turner:1983he,Turner:1985si,DiLuzio:2020wdo}. Consequently, a wide campaign of new experiments is aiming to test the hypothesis that such particles could make up the invisible halo that encloses the Milky Way, see, e.g., \cite{Irastorza:2018dyq}. In the Peccei-Quinn model, the axion is the Goldstone boson of a global $U(1)$ symmetry which is spontaneously broken at the high energy scale $f_a$, and its field forms an extremely cold population of non-relativistic particles whose final density is determined around the QCD phase transition.

In the so called  \emph{post-inflationary scenario}, namely if the symmetry breaking occurs after the end of the inflationary period, the energy distribution of the axion field is highly inhomogeneous at small scales $L_1\sim$ mpc (milliparsec), see Eq.~\eqref{eq:L1}.
As a consequence, most of the axions quickly become  gravitationally bound in small but very dense dark matter halos called axion miniclusters \cite{Hogan:1988mp,Kolb:1994fi}, which would be the smoking gun for axion dark matter in this scenario.
To understand these inhomogeneities, we must follow the physics which gives rise to them.
During the Peccei-Quinn era (temperatures $T\sim f_a$), global cosmic strings arise as a consequence of the symmetry spontaneous breaking of the $U(1)$ symmetry
by the Kibble mechanism \cite{Kibble:1976sj}. 
The axion, as the associated pseudo-Goldstone boson, is effectively a massless particle until the QCD epoch, where it develops a mass through the anomalous breaking of $U(1)$ symmetry by nonperturbative QCD effects (instantons),
which are highly suppressed at high temperatures.
During the range of times where the axion is effectively massless,
the cosmic string network relaxes up to correlation lengths of the order of the horizon in the so-called \emph{scaling} regime.
During this epoch, most of the string energy is converted into axions.%
\footnote{Just a small amount is converted into high-mass particles \cite{Gorghetto:2018myk} and gravitational waves.}
When the QCD potential becomes sizeable (close to the QCD phase transition), cosmic strings are forced to annihilate by domain walls and the population of low-momentum axion fluctuations becomes rapidly non-relativistic and, effectively, a gas of ultra-cold dark matter particles.
Axion fluctuations can be somewhat arbitrarily classified as coming from zero-mode vacuum \emph{realignment} (sometimes called the \emph{misalignment} mechanism) and from the relaxation of topological defects, e.g., strings and walls.
While estimating the dark matter yield from the former is relatively easy, the axion number radiation from strings and walls remains quite uncertain, despite the recent efforts using high-performance numerical simulations \cite{Hiramatsu:2012gg,Fleury:2015aca,Klaer:2017ond,Klaer:2017qhr,Buschmann:2019icd,Gorghetto:2018myk,Gorghetto:2020qws,Buschmann:2021sdq}.  
Having a precise prediction for the final axion dark matter yield as a function of the axion mass can be crucial for its experimental discovery because, by assuming that axions account for the totality of the dark matter, it would allow to pinpoint a unique value of the axion mass to which axion haloscope experiments can be tuned.
Indeed, haloscopes could produce a fast discovery if tuned to the correct mass, but are otherwise forced to perform a slow scan over resonant frequencies, see, e.g., \cite{Irastorza:2018dyq} for a review on axion experiments.  

The key quantity that defines the behaviour of axion emission from strings is the \emph{spectral index} $q$ of the instantaneous emission spectrum.
This quantity describes how the energy, converted from axionic strings into freely propagating axions, is distributed across axion wave lengths.
It is subject to a long-established controversy:
\begin{itemize}
    \item If $q=1$, originally proposed in Refs.~\cite{Harari:1987ht,Hagmann:1990tj} and in agreement with the works of Refs.~\cite{Dine:2020pds,Buschmann:2021sdq}, or in the case where $q\lesssim 1$, as supported by current small string tension simulations \cite{Kawasaki:2014sqa,Gorghetto:2018myk,Vaquero:2018tib,Buschmann:2019icd,Gorghetto:2020qws,Buschmann:2021sdq,OHare:2021zrq}, then the axions from strings do not contribute much to the dark matter and the total axion number density can be computed as an $\mathcal{O}(1)$ modification to the standard misalignment calculation. In this case, the axion mass that gives the dark matter abundance should be found within the range $m_a^{\rm DM}=\mathcal{O}(10-100)~\mu{\rm eV}$. This is also in agreement with simulations of Refs.~\cite{Klaer:2017qhr,Klaer:2017ond}, which introduced a different UV completion of the string cores to effectively achieve physical string tension (see more on this below). 
    \item If $q>1$ and the instantaneous axion spectrum is infrared, as proposed by Refs.~\cite{Davis:1985pt,Davis:1986xc,Battye:1993jv,Battye:1994au} and according to extrapolation of simulations of Refs.~\cite{Gorghetto:2018myk,Gorghetto:2020qws}, the axion number density receives a large enhancement that could substantially exceed the misalignment estimate. In particular, the most recent extrapolated estimate from Ref.~\cite{Gorghetto:2020qws} gives an enhancement  factor of $\sim 200$.%
    \footnote{The estimate of the enhancement factor is $n_a/n_a^{{\rm mis},\theta_i=1}\sim 200$ for a string density parameter $\xi\simeq 15$ at $\log(f_a/H_{\rm QCD})\sim 65$.}
    In this case, the axion mass that gives the observed dark matter abundance is $m_a^{\rm DM}\gtrsim 500~\mu{\rm eV}$. 
\end{itemize}

\noindent\emph{In this work} we want to investigate what each of these scenarios predicts for the distribution and the phenomenology of axion miniclusters.
In particular, we want to compare the predictions of the low-tension simulations of \cite{Vaquero:2018tib,Buschmann:2019icd,Eggemeier:2019khm} with new high-tension simulations following the methodology of \cite{Klaer:2017ond,Klaer:2017qhr} and new simulations of synthetic $q>1$ axion spectra.
The results should give us a good idea of the systematic theoretical uncertainties for minicluster properties and the implications for the post-inflationary scenario of axion dark matter.
The papers is structured as follows:
in Sec.~\ref{sec:piuni} we review the post-inflationary timeline and how we simulate it;
in Sec.~\ref{sec:early} we provide the details of the direct and indirect simulations in the early Universe;
in Sec.~\ref{sec:gravi} we describe their gravitational evolution;
and finally in Secs.~\ref{sec:impl}, \ref{sec:conclusions} we review the implications and conclude.   

\section{The post-inflationary timeline}\label{sec:piuni}

The simulation framework used in this work adopts a mixture of numerical methods, using the publicly available numerical codes \texttt{jaxions}%
\footnote{\href{https://github.com/veintemillas/jaxions}{https://github.com/veintemillas/jaxions}}
and \texttt{gadget-4}~\cite{Springel:2020plp}%
\footnote{\href{https://wwwmpa.mpa-garching.mpg.de/gadget4/}{https://wwwmpa.mpa-garching.mpg.de/gadget4/}}
as first described in Ref.~\cite{Eggemeier:2022hqa}.
For each simulation we \emph{continuously} evolve a realistic field configuration from a typical redshift $z_i\sim 10^{13}$ to a final redshift $z_f\sim 10^2$ in several steps: 

\begin{enumerate}
    \item \emph{Scaling and QCD phase transition}. From an initial redshift $\sim z_i$ we solve the relativistic Klein-Gordon equation on the lattice with a typical resolution of $\mathcal{O}(10^{10})$ grid sites.
    We make a distinction between two different methods (described in detail in Sec.~\ref{sec:early}): 
    \begin{itemize}
    \item In \emph{direct} simulations, we solve the equation for complex fields $\phi$ with a string-wall network, the axion being an angular field.
    We can do this with two different theories: with a single complex scalar as in \cite{Vaquero:2018tib}, which produces low-tension strings,
    or with the effective high-tension theory of Refs.~\cite{Klaer:2017ond,Klaer:2017qhr}, which includes two complex scalars and a vector field to create a network of high-tension strings, plus extra heavy degrees of freedom.
    We evolve the fields until strings and walls have completely disappeared ($z_i/2\sim z_i/4$ depending on the string tension) and then continue with axion-only simulations. 
    \item In \emph{indirect} simulations we calculate the initial axion spectrum from strings analytically at $z_i$.
    We then create a synthetic axion background where we enforce this axion spectrum, without any phase coherence strings or walls, and evolve this axion field through the QCD epoch until it becomes non-relativistic.
    \end{itemize}
    We stop the simulations at $\sim z_i/5$ when the axion field contains mostly non-relativistic and small-amplitude oscillations with very occasional nonlinear collapse events.
    These collapse events
    (called \emph{axitons} in this context~\cite{Kolb:1993hw})
    become rarer with time, and the length scale on which they occur becomes shorter as the axion mass increases.
    At $z \sim z_i/5$ our lattice can no longer resolve them.

    Given that from $\sim z_i/5$ the total axion number density is comovingly conserved (neglecting axitons), this is the only step required to predict the axion dark matter mass.
    The remaining steps are needed to determine the spatial inhomogeneity of the axions and the presence of miniclusters.
    \item \emph{WKB}. From $\sim z_i/5$ to $\sim z_i/20$ we evolve the fields analytically in Fourier space by neglecting the non-linearities of the QCD potential in the WKB approximation as in \cite{Vaquero:2018tib} to get rid of the axitons. This evolution conserves axion number and the density fluctuations at large and intermediate scales, correctly accounting for the dispersion of axion waves on those scales. It only fails at representing the fate of axions in axitons, which have become very rare and occur in small, spatially localised regions. 
    \item \emph{Free-streaming}. From $\sim z_i/20$ to $z\sim 10^5$ we solve the non-relativistic Schr\"odinger-Poisson system of equations on a reduced lattice. We neglect axion self-interactions, which are negligible after the WKB period, but we include gravity. We stop the simulation when we reach the maximum phase gradient that can be resolved,\footnote{The gradient $\nabla S$, with $S$ the complex's scalar phase, is related to the local differences in the gravitational potential.} see ~\cite{Eggemeier:2022hqa}. This allows us to keep track of the small effects of gravity in the linear regime while accounting for the dispersion of axion waves, not altogether negligible over several orders of magnitude in redshift.
    \item \emph{Gravitational collapse}. From $z\sim 10^6$ to $z_f$ we solve the Vlasov-Poisson system with cold N-body particles. The system can be evolved with a collisionless system of equations since any wave-like feature (i.e., oscillatory modes that oppose the gravitational force) appears at spatial scales $\ell\lesssim \lambda_J$, which happen to be smaller than the discretisation scale of our simulations. The final limitation on the timeline imposed by these simulations stems from the smallness of the simulation box in physical units, since box-size ($\sim$ pc) scales are non-linear around a typical redshift of $z\sim 100$. As done in Refs.~\cite{Eggemeier:2019khm,Eggemeier:2022hqa}, we fix $z_f$ in each simulation by looking at the density fluctuations in the axion energy density.
\end{enumerate} For redshifts $z<z_f$ our simulations are unreliable, therefore we cannot numerically evaluate the distribution and shape of miniclusters at $z=0$. Moreover, although miniclusters can have densities up to $\rho\sim 10^{10}~M_{\odot}~{\rm pc}^{-3}$, with $M_{\odot}$ the solar mass, they can be disrupted by interactions with forming stars and tidal forces of the forming galaxy, see e.g.~\cite{Dokuchaev,Tinyakov:2015cgg,Kavanagh:2020gcy,Dandoy:2022prp,Shen:2022ltx}, which has critical implications for both direct and indirect detection (see Sec.~\ref{sec:impl}).

\subsection{Axion dark matter mass and domain wall number}

The different approaches that we use for simulating the cosmology of post-inflationary axion dark matter predict different dark matter abundance functions $\Omega_a=\Omega_a(m_a)$ and thus a substantially different axion dark matter mass. The direct simulation with a single scalar field as in Ref.~\cite{Vaquero:2018tib} suggests $m_a\sim 17~\mu{\rm eV}$, while the high-tension gauged double-complex-scalar of Klaer and Moore \cite{Klaer:2017ond} requires $m_a\simeq 26~\mu{\rm eV}$. Indirect methods, which calculate the axion spectrum from strings by extrapolating the instantaneous emission spectrum, lead to $m_a\gtrsim 500~\mu{\rm eV}$ for \cite{Gorghetto:2020qws} and $m_a \sim 110\pm 70~\mu{\rm eV}$ for \cite{Buschmann:2021sdq}. On the other hand, all simulations are for the most part insensitive to the value of $m_a$ because it can be rescaled away, up to negligible terms.
The dependence enters only into the results when we translate code units to physical units.
Indeed, the characteristic length scale, given by the moment when the Hubble expansion rate equals the axion mass, $m_a(t_1)=H(t_1)$, turns out to be quite insensitive to the axion mass. Numerically one finds \cite{Vaquero:2018tib}
\begin{equation}
    L_1=\frac{1+z(t_1)}{H(t_1)}\simeq 0.0362~{\rm pc}\left(\frac{50~\mu{\rm eV}}{m_a}\right)^{0.167}, \label{eq:L1}
\end{equation} 
where the numerical mass dependence is based on lattice QCD results for axion cosmology~\cite{Borsanyi:2016ksw}.
Therefore, the characteristic mass $M_1$ contained within a volume $L_1^3$ depends on the axion mass as $m_a^{-1/2}$.%
\footnote{For instance, taking $m_a=0.5$ meV as motivated by indirect simulations with $q=2$ and $N_{\rm DW}=1$~\cite{Gorghetto:2020qws}, the $M_1$ definition changes by a factor $\sqrt{10}\simeq 3.16$ with respect to benchark axion mass $m_a=50~\mu{\rm eV}$.} As we will see, this implies that the variance in the results between the different techniques is larger than the differences induced by using the appropriate axion dark matter mass for each method. Therefore, in the following we will mostly compare the results in code units, keeping in mind that the physical scales might have some changes due this normalisation. 

In addition, in this paper we focus on axion models with domain wall number $N_{\rm DW}=1$, where only one physically CP conserving vacuum of the axion field exists, $a=0$. With this choice, strings are destroyed by domain walls without requiring model-dependent explicit PQ breaking terms \cite{Zeldovich:1974uw,Sikivie:1982qv}. 

\section{Early Universe dynamics}\label{sec:early}

The first step of the timeline involves solving the relativistic Klein-Gordon equation for the axion degree of freedom in the early Universe. For the sake of minimality, we assume that the string networks evolve during radiation domination.

Ideally, one would like to simulate from the moment when the global symmetry spontaneously breaks until after the string network breaks up and axion fluctuations become nonrelativistic.
In practice this is impossible because of the severe hierarchy of scales between the axion-string core size, which must be at least inverse-GeV scale and is more likely $\sim 1/f_a \sim 1/(10^{11}\,\mathrm{GeV})$, and the inverse Hubble scale at the QCD epoch, which is of order meters.
To successfully simulate both scales would require a resolution of at least $r_{\mathrm{max}}/r_{\mathrm{min}} \sim f_a/H \sim 10^{30}$.
A lattice with $10^{30}$ grid points on a side is clearly unfeasible.
Currently, the largest grids attained in simulations with large computer clusters have managed $\sim (10^4)^3$ Cartesian grids~\cite{Vaquero:2018tib}, or an equivalent to $\sim (10^5)^3$ with adaptive mesh-refinement around the strings \cite{Buschmann:2021sdq}. 
One might hope that the short-distance physics is not important and that simulations which treat the scale ratio as only $\sim 10^3$ rather than $\sim 10^{30}$ would capture the right physics.
But this is unclear, as the global strings store most of their energy as gradients of the axion field $\nabla a \propto {\rm (distance)^{-1}}$, and hence the logarithm of the large scale hierarchy appears in the string tension,
$\mu={\rm  Energy/Length}\sim \pi f_a^2 \log(r_{\rm max}/r_{\rm min})$.
Therefore, to simulate the correct string tension, i.e., the correct dressing parameter $\kappa = \mu/(\pi f_a^2) = \log(r_{\rm max}/r_{\rm min})\sim \log(f_a/H)\sim 70$ at the QCD phase transition, we need a dynamical range of $r_{\rm max}/r_{\rm min}\sim f_a/H\sim 10^{30}$, \textsl{or} we need to extrapolate to this value, \textsl{or} we need an effective description which captures this large string dressing parameter.

In order to overcome these limitations, two main approaches have been taken in the literature, which we label in the following as {\rm direct} and {\rm indirect} methods.  

\subsection{Direct simulations}

In direct simulations one aims to keep track of the full system of equations of motion for the axion field around the characteristic time $t_1$. This was the approach taken in the simulations of Refs.~\cite{Vaquero:2018tib,Kawasaki:2014sqa,Fleury:2015aca,Buschmann:2019icd,OHare:2021zrq}, which follows the evolution of a complex scalar field, $\phi (x)$, whose angular component is the axion, $\phi (x)=r(x)  e^{i a(x)/f_a}$, and whose radial mode provides the UV regulator of the string density, $r_{\rm min}\sim m_r^{-1}$, with $m_r$ the inverse mass of the radial mode.
Such simulations, which include contributions from domain walls and non-linear dynamics, can only be done for unnaturally small values of $m_r\ll f_a$ (for which indeed the radial mode, sometimes called \emph{saxion}, is actually experimentally excluded), reaching a string tension parameter $\kappa = \log(m_r/H) \sim 9$. To overcome this technical problem, the authors of Ref.~\cite{Klaer:2017qhr} developed a numerical approach to effectively simulate a system of strings with arbitrarily larger tension, by simply adding new degrees of freedom in the UV theory and without having to employ extremely large lattices. Indeed, in this method the axion string tension becomes \emph{effectively tunable}, since it is simply set by a function of $\mathcal{O}(1)$ discrete charges. We briefly mention the relevant aspects of this method here, for details see \cite{Klaer:2017qhr}.

The theory is composed of one $U(1)$ gauge field $A_{\mu}$ and two complex scalar fields $\phi_{1,2}=\vert\phi_{1,2}\vert e^{i\theta_{1,2}}$ with charges $q_1,q_2$ under the gauge field.
A linear combination of the complex fields' phases $q_1 \theta_1 + q_2 \theta_2$ is ``eaten'' via the Higgs mechanism, making the gauge field massive.
The orthogonal combination $a \propto (q_2 \theta_1 - q_1 \theta_2)$ is the axion field.
Provided that the initial conditions place $\phi_1$-strings and $\phi_2$-strings in the same positions, the cosmic string then has a global axionic charge, but an abelian-Higgs core which dominates the string tension.
The string tension is approximately $\mu \simeq \pi(v_1^2+v_2^2)$ (where $v_{1,2}$ are the VEVs of the two scalars), but the value of $f_a$ is much smaller,
\begin{equation}
        f^2_a = \frac{v_1^2v_2^2}{q^2_1v^2_1+q^2_2v^2_2}, 
\end{equation} 
which can be engineered to be parametrically small by adjusting the $q_{1,2}$ charges and $v_1/v_2$ so that the effective tension $\kappa=\mu/\pi f_a^2$ is as large as needed, 
\begin{equation}\label{eq:tunable}
\kappa \simeq \frac{\pi(v_1^2+v_2^2)}{\pi f^2_a}=\frac{(v^2_1+v^2_2)(q^2_1v^2_1+q^2_2v^2_2)}{v^2_1v^2_2}=2(q^2_1+q^2_2), \end{equation}
where in the last step we have taken $v_1=v_2$.

In this work we use the numerical code of Refs.~\cite{Klaer:2017qhr,Klaer:2017ond} to simulate this system and create an interface to \texttt{jaxions} to continue the evolution once the strings and walls disappear.
In our runs we simulate with a resolution up to $N^3=4032^3$ lattice points for physical sizes $L/L_1=2,3.6$, and we use a string core thickness $m_rL/N=1$.%
\footnote{Convergence tests ~\cite{Vaquero:2018tib,Gorghetto:2018myk} show that string core resolution parameter $m_rL/N$ has to be within the value 1.5.}
We simulate from an initial redshift $z_i=10^3z_1\sim 10^{16}$ until $z_1/4$, by which time the string network has been annihilated.

The physical size $L/L_1$ is limited for several reasons, such as the fact that in simulations with the correct $\kappa\sim 60-70$, the string tension delays the destructive effects of domain walls, the axion mass is a fast growing function ($m_a \propto (z_1/z)^{n/2}$, with $n\sim 7$) and the UV degrees of freedom (radial and gauge modes) have to have reasonably larger masses than the axion to avoid unphysical $2\to 1$ axion-to-heavy-mode processes.
If we can afford $N^3$ grids due to RAM or computational resources, and choose $m_rL/N=1$, which is sufficient to resolve the radial modes around the string cores,%
\footnote{The code uses the PRS trick~\cite{Press:1972zz,Press:1973iz}, where the physical core width increases with time to keep the UV physics at a fixed comoving scale and therefore a fixed grid scale.}
we have that $m_a/m_r = z_1(z_1/z)^{n/2}L/N/L_1$.
Since the destruction of the network happens for $z_1/z\sim 4$ for the relevant values of $\kappa$, we find that limiting $(m_a/m_r)<0.45$ until $z_1/z\sim 4$ requires $L/L_1<(z/z_1)^{n/2}N\sim 3.9 (N/4000)$.

Once strings and walls have annihilated, we switch to axion-only simulations with the \texttt{jaxions} code. We calculate the axion field following \cite{Klaer:2017ond,Klaer:2017qhr} and evolve it with the non-linear Sine-Gordon equation as in \cite{Vaquero:2018tib} until axitons cannot be properly resolved, $m_a L/N\sim O(1)$, which happens to be only $\sim z_1/5$. 

We show the power-spectrum of density fluctuations from our fiducial simulation $L/L_1=3.6,~q_1=3,~q_2=4,~\kappa=60$ in Fig.~\ref{fig:spectra} as an orange line. Compared with the result of low-tension simulations of \cite{Vaquero:2018tib}, the peak moves to higher $k$ by a  factor of 10, the UV tail is smoother and the peak is lower.
The shift to the UV is not entirely unexpected, as high-$\kappa$ networks are expected to be denser, and the larger string tension means that the action of domain walls destroys the network when the axion mass is larger, therefore producing smaller string loops which produce heavier axions with shorter propagation distances.
Very roughly, one expects $m_a$ at network breakup to scale with $\kappa H$, so a factor-10 increase in $\kappa$ leads to nearly a factor-10 increase in the axion mass at string-network breakup and a factor-10 reduction in the final structure size.

\subsection{Indirect simulations}

This method, adopted in Refs.~\cite{Gorghetto:2018myk,Gorghetto:2020qws,Buschmann:2021sdq,RS}, consists of two steps. In the first step, one tries to build an educated guess of the axion spectrum at the QCD phase transition (redshifts of $z_1$) by identifying the key properties of the string network evolution with small-tension simulations and extrapolation to the physically relevant values. In the second step, one runs a simulation of the non-linear evolution during the QCD phase transition of an axion field built from such a spectrum. 

Simulations of the string network with the single complex scalar theory show a simple almost scale-invariant solution with small logarithmic deviations \cite{Fleury:2015aca,Gorghetto:2018myk,Vaquero:2018tib,Gorghetto:2020qws,Buschmann:2021sdq}, and Refs.~\cite{Gorghetto:2018myk,Gorghetto:2020qws} argue that the system has indeed an attractor. By studying the properties around the attractor, one gets rid of the arbitrariness of the initial conditions in the simulations and would find the correct trends that can be safely extrapolated. 

The string network loses energy density into axions at a rate
\begin{equation}
    \Gamma_a=\frac{1}{R^4}\frac{\partial}{\partial t}(R^4\rho_{\rm str})  
    \xrightarrow{\kappa \gg 1} \xi \mu (2 H)^3, 
\end{equation} 
 where $R$ is the scale factor, $\rho_{\rm str}=\xi \mu/t^2$ is the energy density of the string network,  $\xi$ is the $\mathcal{O}(1)$ parameter representing the string network density (string length per Hubble volume measured in Hubble lengths~\cite{Gorghetto:2018myk}) and the large-tension limit is taken from \cite{Gorghetto:2018myk}. 
It is then useful to define a differential spectral transfer rate $\partial \Gamma_a/\partial k\propto F(k,m_s)$ at each instant giving the rate of production of axions of momentum $k$. 
The \emph{instantaneous emission spectrum} $F$ peaks around an IR cut-off $k\sim H$ and shows a power-law decrease towards large $k$ and finally has a UV cut-off at the radial mode mass $m_r$ \cite{Gorghetto:2018myk}.  
By understanding its behaviour as a function of $\kappa$ we can extrapolate to the physical values and integrate to obtain the axion energy spectrum, 
\begin{align}
    \frac{\partial\rho_a}{\partial k}&=\int^{t}{\rm d}t\frac{\Gamma'_a}{H'}\left(\frac{R'}{R}\right)^3 F\left(\frac{k R}{H'R'},\frac{m'_s}{H'}\right) \label{drho_dk_from_F}\\
    &=\frac{k^2}{(2\pi)^3L^3}\int {\rm d}\Omega_k\vert \dot{a}_k\vert^2,
\end{align} 
where in the second line we have expressed the axion energy density as kinetic energy, $k=\vert{\bf k}\vert$ and $\Omega_k$ is the solid angle. 
The value of $\xi$ has been found to slowly approach the attractor, $\xi=\xi_0+ 0.24(2)\kappa$ \cite{Gorghetto:2020qws} with compatible values found in \cite{Klaer:2017ond,Vaquero:2018tib,Buschmann:2021sdq}. 

\begin{figure}
    \centering
    \includegraphics[width=0.55\textwidth]{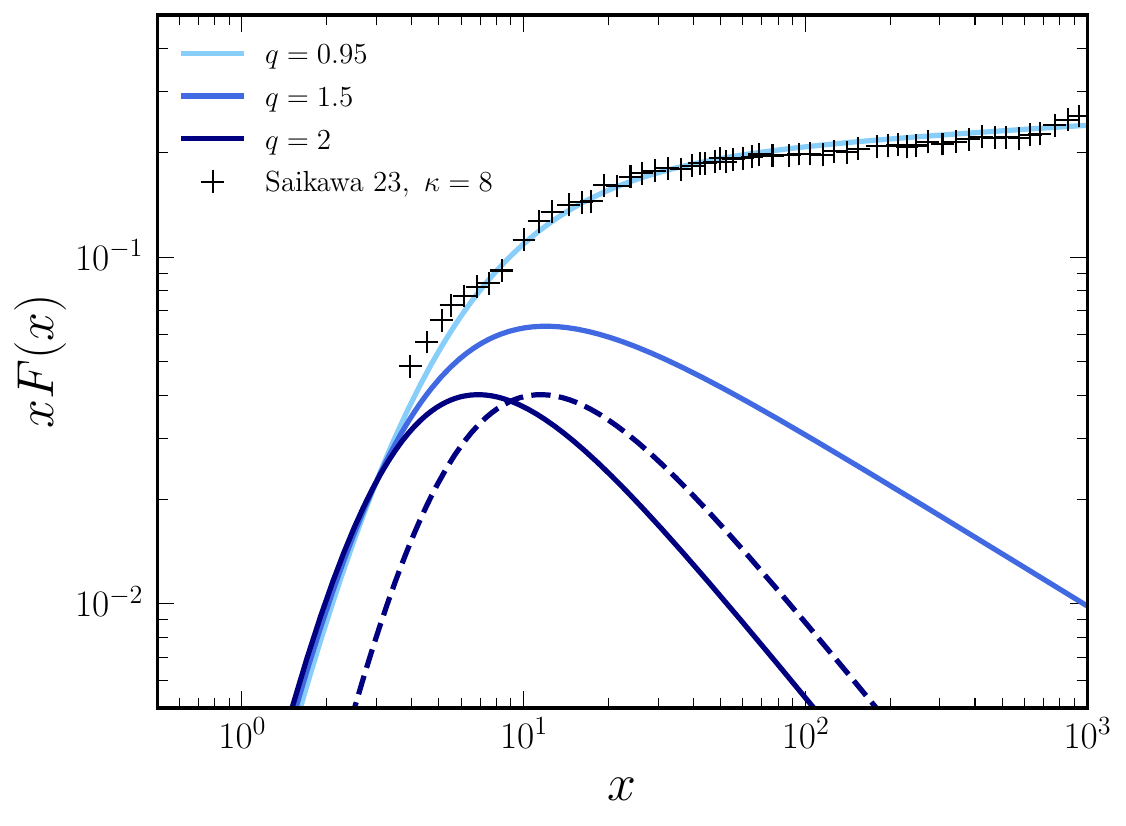}
    \caption{Normalised instantaneous emission spectra for different values of $q$, calibrated from results of \cite{RS} (Saikawa 23) at $\kappa=\log(m_s/H)=8$ and where $x=k/H$. The dashed line shows a spectrum with $q=2$ and $x_0=5$ that we use in our simulations. }
    \label{fig:F}
\end{figure}

Recently, the calculation of the instantaneous emission spectrum has been refined in Ref.~\cite{RS} by developing a new method to extract the characteristic features of the spectrum and suppress possible contaminations from field fluctuations. With the refined method, the evolution of the instantaneous emission spectrum has been analyzed up to $\kappa=\log(m_s/H) \approx 9.08$. In this work, we use those results to calibrate the analytical function of $F(x,y)$ (see Fig.~\ref{fig:F}), which is used to produce initial conditions for our simulations solving only the axion field. 

We parameterise the instantaneous emission spectrum as
\begin{equation}
    F(x;x_0,q,\epsilon)= F_0 \frac{(x/x_0)^3}{\left(1+(\frac{x}{x_0})^{\frac{3+q}{\epsilon}}\right)^{\epsilon}}, \quad x=k/H, \quad x < m_r/H, 
\end{equation} 
with $x_0$ indicating the IR cutoff, $\epsilon$ the broadness of the peak and $q$ the spectral index of the main power law. The spectrum is cut off sharply at the radial mass. The normalisation $F_0$ is such that the $k$ integral is normalised to 1 as required by \eqref{drho_dk_from_F}.

The value of $q$ is the most relevant for the final axion abundance. 
The recent and thorough study of \cite{Gorghetto:2020qws} finds that the best attractor extrapolation seems to be a linear dependence $q\simeq 0.49+0.056 \kappa$, that gives $q(\kappa=60)\sim 5$. The unique AMR-boosted simulation of \cite{Buschmann:2021sdq} finds a best fit of $q\simeq 1.02(3)$ with no significant upwards trend. 
According to \cite{RS}, the discrepancy of the above references can be traced to their different initial conditions, discretisation effects and analysis choices. We find that both a linear trend $q=q_0+q_1\kappa$ and a saturating $q=q_0-q_1/\kappa^2$ fit the data very well.
No significant preference for one or the other can be drawn from the analysis of \cite{RS}. 

We note however that if $q\simeq 1$, then the spectrum will be close to that of the direct simulations, which already reaches $q\sim 1$.
Therefore, we choose to investigate the values $q>1$, which lead to completely different initial conditions. One can show that any value of $q>1$ leads to an energy spectrum $\partial\rho/\partial k \sim 1/k$ up to small corrections with an 
absolute normalisation that only weakly depends on $q$ \cite{Gorghetto:2018myk,Gorghetto:2020qws}, so we consider only the case $q=2$ for simplicity. The difference in predictions between $q=2$ and $q=5$ are much smaller than the differences between direct and indirect simulations. 

For $q>1$, the axion density is so large that the expectation value of the axion is larger than $a/f_a\sim \pi$, i.e., the field wraps frequently around the mexican hat potential. Moreover, the energy density is so large that we can neglect the standard misalignment and even the few cosmic strings present at $z_1$, which will be destroyed by domain walls by $z_1/4$ \cite{Gorghetto:2020qws}. This is convenient, since we cannot simulate these strings with the correct value of $\kappa$. Moreover, neglecting the strings allows us to simulate the axion field only, which automatically gets rid of the problem of axion energy leaking into radial modes.

\exclude{
For our fiducial simulation we use $q=2$, $\epsilon =3$, a density-sensitive IR cut-off 
\begin{equation}
    x_{\rm peak} = 1.86 + 2.36(10)\xi^{0.52(6)}
    \quad, \quad 
    x_0 = x_{\rm peak} \left(\frac{q}{3}\right)^\frac{\epsilon}{3+q}, 
\end{equation} 
as found in \cite{RS} and $\xi=-1.97(14)+0.26(1)\kappa$ (from \cite{RS}, compatible with \cite{Gorghetto:2020qws}). 
Note that the extrapolated value of $x_{\rm peak}$ is substantially larger than the measured value of the simulations at small string tension, which reaches $x_{\rm peak}(\kappa\sim 8)\simeq 4$, as seen in Fig.~\ref{fig:F}. 
}

The time-integrated spectrum is very insensitive to the $\kappa$-dependence of $\xi,~x_0$ and $q$, mostly depending on the final values of the parameters, so we prepare our fiducial spectrum simulation with
$\xi=15$,
$\epsilon = 3$,%
\footnote{We also observed a trend on $\epsilon$ that increases with $\kappa$, implying that the IR peak could be much broader. Such an effect of the broadening of the IR peak should be kept in mind as a potential ambiguity, though understanding of the physical origin of the trend is beyond the scope of this work.}
$x_0=10$, and
$q=2$
at $z_1$ assuming $\kappa(z_1)=60$.

We run our lattice simulations using the {\tt jaxions} code and with initial conditions set by the $\partial\rho_a/\partial k$ spectrum, obtained by integrating our parametrisation of the instantaneous spectrum.%
\footnote{We choose the Fourier modes of the axion field and its derivatives with amplitudes sampled from Gaussian distributions with means $a(\vec k) \propto \sqrt{k^{-2}\partial\rho/\partial k}$ and $\dot a(\vec k) \propto \sqrt{\partial\rho/\partial k}$ and random phases. The procedure is analogous to \cite{Gorghetto:2020qws}, except there a flat distribution is used instead of a Gaussian.} 
We use $4096^3$ grids and physical length $L/L_1=1,3$. We start our simulations at $z_1$ and run until axitons cannot be resolved ($z_1/6$ for $L/L_1=1$ simulations). 

The non-linear evolution of large-amplitude axion fields during the QCD phase transition has been thoroughly examined in \cite{Gorghetto:2020qws}.
The most important aspect for this work is that the nonlinearities destroy low-$k$ axions, shifting the spectrum peak towards larger $k$.
This dynamics makes the final spectrum largely insensitive to the details of $x_0, \epsilon$, as well as to the inclusion or not of the final cosmic strings and their domain walls. 

We show the power-spectrum of density fluctuations from our fiducial simulation in Fig.~\ref{fig:spectra} as a blue line. We notice how non-linearities have shifted the spectrum peak from $k/k_1\sim 10$ to a value $k/k_1\sim 400$, much higher than even the high-$\kappa$ simulations. The peak height is comparable to high-$\kappa$ simulations and about half that in low-$\kappa$ simulations. 

\subsection{Density fluctuations}

\begin{figure}
    \centering
    \includegraphics[width=0.55\textwidth]{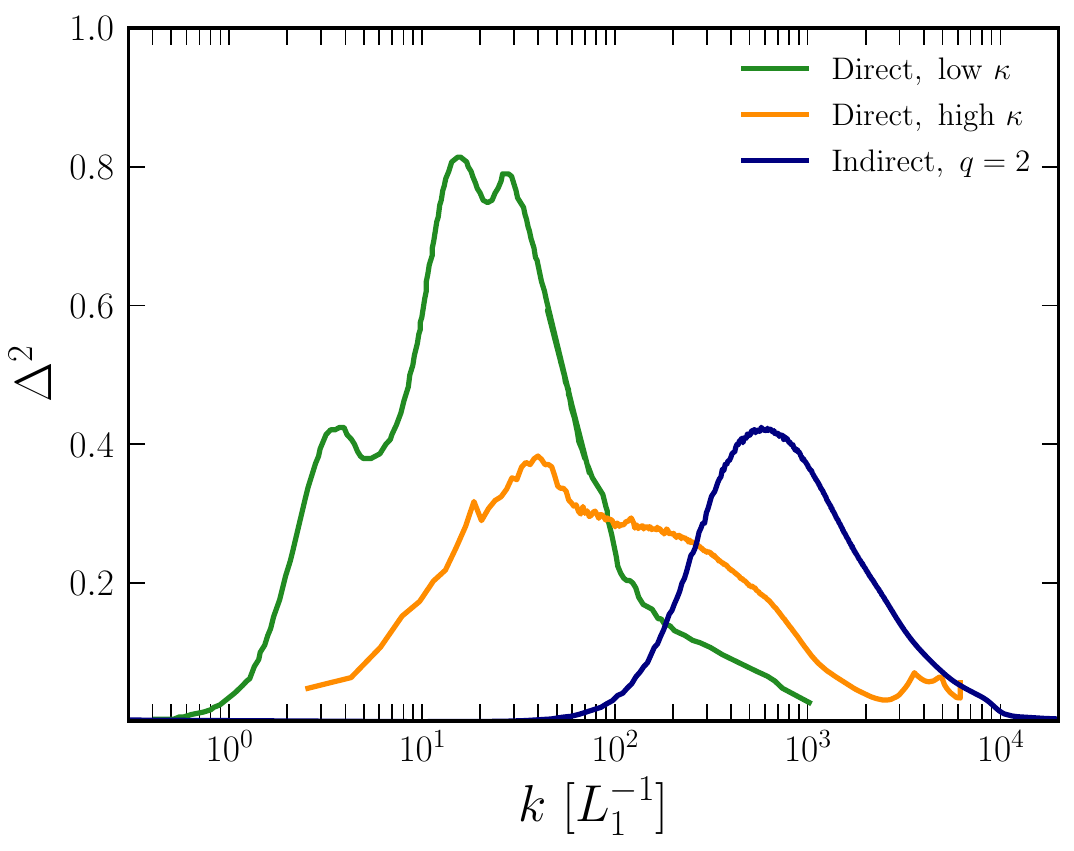}
    \caption{Axion energy density dimensionless power spectrum for direct and indirect simulations as described in Sec.~\ref{sec:early} and compared to small string tension results of Ref.~\cite{Vaquero:2018tib}.}
    \label{fig:spectra}
\end{figure}

To study the fluctuations in the axion energy density, we define the local density contrast
\begin{equation}
    \delta_a({\bf x})=\frac{\rho_a({\bf x})-\langle\rho_a\rangle}{\langle\rho_a\rangle},
\end{equation}
where $\langle\rho_a\rangle$ is the spatially averaged energy density. By taking the Fourier transform of $\delta_a$, $\tilde{\delta}_a({\bf k})={\cal F}[\delta_a({\bf x})]$, we can build the dimensionless power spectrum in a comoving volume $V$ as
\begin{equation}
    \Delta^2_a(k)=\frac{k^3}{2\pi^2V}\langle\vert\tilde{\delta}_a({\bf k})\vert^2\rangle_{\vert{\bf k}\vert=k},
\end{equation}
depicted in Fig.~\ref{fig:spectra}. On large scales $k\lesssim 1/L_1$, the spectrum should represent a system of uncorrelated patches, described by a white-noise law $\Delta^2_a\propto k^3$ and confirmed by direct simulations at small $\kappa$, see, e.g., Ref.~\cite{Vaquero:2018tib}. This behaviour is analogously seen in the indirect simulations, where at small $k$ the scale dependence can be fitted by
\begin{equation}
    \Delta^2_{a}(k)=3.8\times 10^{-8}(kL_1)^3,~~~~~({\rm Indirect},~q=2).
\end{equation}
Compared to results of Ref.~\cite{Vaquero:2018tib}, the position of the peak is shifted by more than one order of magnitude towards smaller spatial scales, leading to much suppressed amplitude on large scales. This effect is related to the initial spectrum with $q>1$ and the fact that nonlinear dynamics caused by the axion potential are delayed by a factor $\propto \sqrt{\xi\kappa}$~\cite{Gorghetto:2020qws}. The direct simulations with large $\kappa$ have much more power at small scales. A fit forcing $\Delta\propto k^3$ gives\footnote{In this case, the volume is similar to the correlation length. We cannot see much of the white-noise regime and we must force the exponent to find this result.}
\begin{equation}
    \Delta^2_{a}(k)=1.3\times 10^{-3}(kL_1)^3,~~~~~({\rm Direct,~high}~\kappa).
\end{equation}
The non-trivial shape in the dimensionless spectrum, such as the presence of small oscillatory features, seems to be peculiar of the direct simulation method that includes the contribution of forming domain walls and collapsing strings. For this reason, the exact shape of the spectrum for the indirect simulation will need to be revisited in future works.    

\section{Gravitational dynamics}\label{sec:gravi}

Density fluctuations collapse into miniclusters at $z\gtrsim z_{\rm eq}$, with $z_{\rm eq}\sim 3400$ the redshift of matter-radiation equality, and effectively decouple (locally) from the Hubble flow. A non-linear model for the collapse of overdensities has been first studied in Ref.~\cite{Kolb:1994fi}, and recently applied by Refs.~\cite{Enander:2017ogx,Ellis:2020gtq,Eroncel:2022efc,Chatrchyan:2023cmz}. Here, we go beyond this treatment and solve the collapse of overdensities into miniclusters numerically, using a modified version of the code \texttt{gadget-4}, by allowing us to evolve during radiation domination and across matter-radiation equality. We map the field configuration by assigning one particle per grid site and choosing a particle mass based on the energy density $m_{i}({\bf x})\propto \delta_a({\bf x})$, as introduced in Ref.~\cite{Eggemeier:2022hqa}. The typical particle mass is $m_{\rm av}\simeq 10^{-18}~M_{\odot}$. We simulate $512^3$ particles using the TreePM method for the particle force, on a $512^3$ mesh and with softening lengths $\ell_{\rm soft}=\Delta/60\sim 1$ AU, where $\Delta=L_{\rm box}/512$ is the average inter-particle distance. We adopt the timestep ${\rm d}t=\sqrt{2\eta \ell_{\rm soft}/a_x}$, where $a_x$ is the particle acceleration, with accuracy set by $\eta=0.01$.

\subsection{Energy budget}

\begin{figure}
    \centering
    \includegraphics[width=0.48\textwidth]{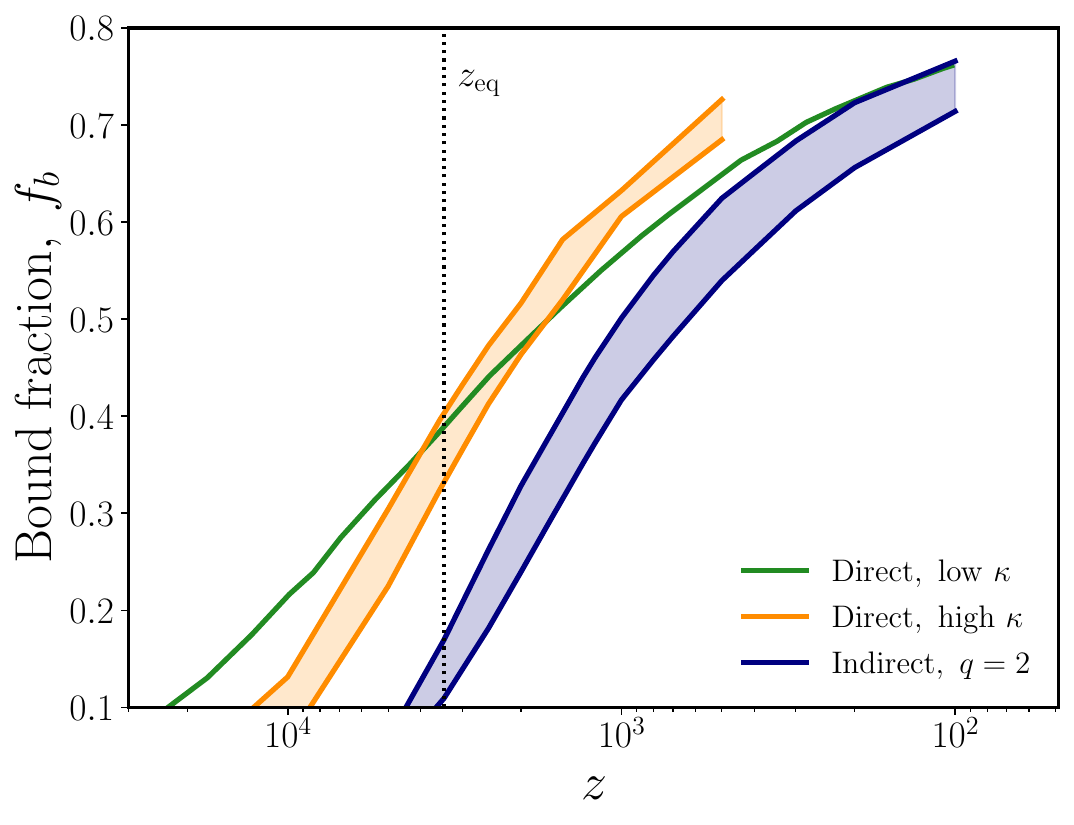}
    \includegraphics[width=0.48\textwidth]{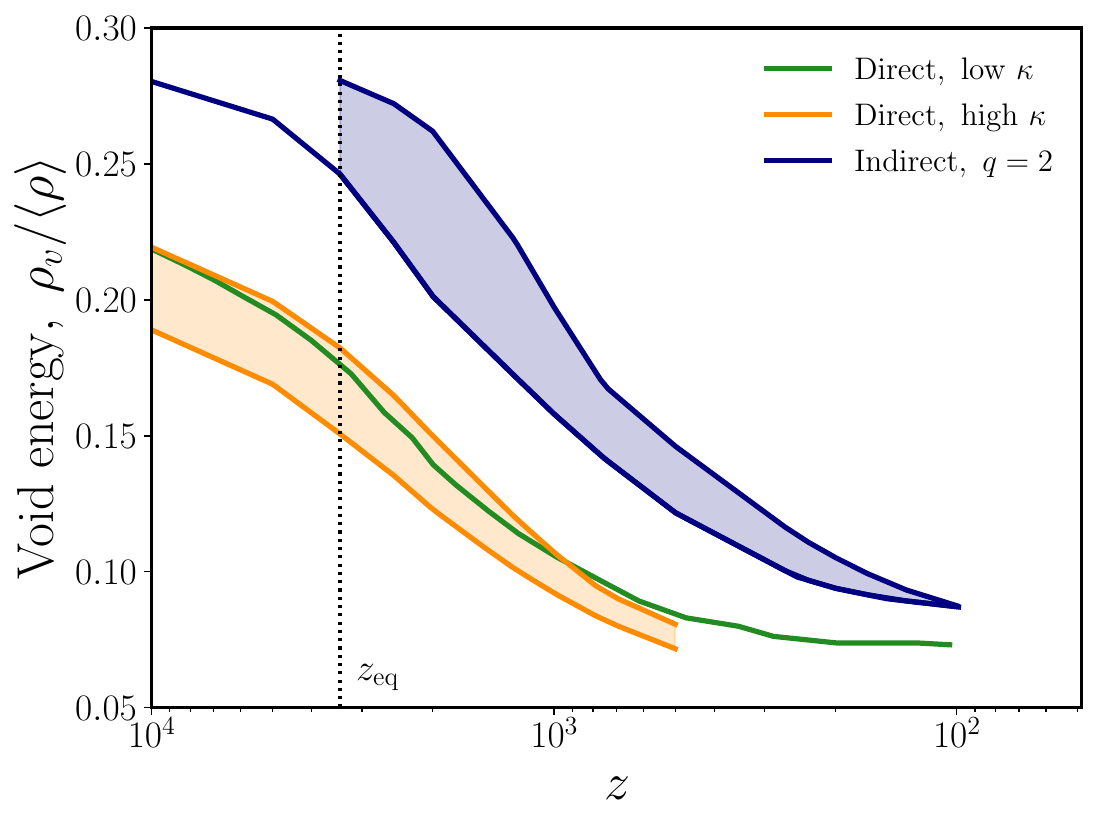}
    \caption{\emph{Left}: Bound fraction as defined in Eq. \eqref{eq:boundf} for direct and indirect simulations and compared to the results of Ref.~\cite{Eggemeier:2019khm}. The colored bands represent the uncertainty over the different simulations. \emph{Right}: Void average density, normalised by the overall average energy density, as defined in Eq. \eqref{eq:boundf} for direct and indirect simulations and compared to the results of Ref.~\cite{Eggemeier:2022hqa}.}
    \label{fig:budget}
\end{figure}

Most of the axions are fated to become gravitationally bound in the miniclusters, as already seen in Refs.~\cite{Eggemeier:2019khm,Eggemeier:2022hqa} and quantified by defining a bound fraction
\begin{equation}
    f_b=\frac{\sum_{j} m_j}{M_{\rm tot}},~~~j\in\{{\rm Bound~particles}\},\label{eq:boundf}
\end{equation}
where $M_{\rm tot}$ is the total sum of the particles masses. The bound particles are taken from the output \emph{friends-of-friends} halo finder, which identifies groups of at least 32 particles. We have checked that by calculating $f_b$ using the \emph{subfind} catalogue instead, there is at most a 5\% difference at the final simulation time. In the left panel of Fig.~\ref{fig:budget} we show the bound fraction as a function of redshift $z$ and compare it with the low $\kappa$ result of Ref.~\cite{Eggemeier:2019khm}. We notice that the results for the $q=2$ indirect simulations and the results from Ref.~\cite{Eggemeier:2019khm} become comparable with $f_b\sim 0.75$ at $z=99$, despite the delayed minicluster formation of the former. Direct high-$\kappa$ runs seem to also stabilise to a result close to $f_b=0.7$ at the final simulation time $z\sim 500$. For our new simulations we notice a common steepness in the growth of $f_b$ with time, as highlighted in Fig.~\ref{fig:budget}. 

While the energy budget is dominated by miniclusters and their halos, most of the volume is occupied by \emph{minivoids}, making up around 70-75\% of the simulation volume~\cite{Eggemeier:2022hqa}. Depending on the subsequent fate of the miniclusters, if large areas ($\sim L_1$) of our Universe are substantially underdense, the expected detection signal of \emph{all} axion haloscopes would be reduced by a factor $\propto \rho_a^{1/2}$. Our goal is therefore to estimate the typical value of the axion energy density in underdense regions, as done for the first time in Ref.~\cite{Eggemeier:2022hqa}. We adapt the same methodology as the latter, which identifies voids of size $\ell\in [5,50]\Delta\sim [1,10]$ mpc with a fiducial density threshold $\delta_a^{\rm thr}=-0.7$%
\footnote{The fiducial value is motivated by convergence tests done in Ref.~\cite{Eggemeier:2022hqa}.}
and finds the following average void densities at the final simulation time: 
\begin{align}
    &\rho_{\rm v}/\langle\rho\rangle\simeq 0.07-0.08 ~~~&({\rm Direct~high~}\kappa,~z=499),\\
    &\rho_{\rm v}/\langle\rho\rangle\simeq 0.09, ~~~&({\rm Indirect},~q=2, z=99). \label{eq:void}
\end{align}
Similarly to the case of the bound fraction, we notice how all methods seem to converge to a similar void density value of about $8\%$ of the overall average density. This is also in agreement with simulations at low $\kappa$~\cite{Eggemeier:2022hqa}, as depicted in Fig.~\ref{fig:budget}.

\subsection{Minicluster mass distribution}

To study the mass distribution of miniclusters we parametrise the \emph{halo-mass function} (HMF) as the number density of miniclusters of mass in the interval $(M,M+{\rm d}M)$. This can be calculated directly from our numerical N-body simulations and can be estimated analytically under some approximations. 

\subsubsection{Analytical expectations}
In the Press–Schechter approach~\cite{Press:1973iz}, one associates halos of a given mass $M$ to a radius $R_M\simeq(3M/4\pi\langle\rho\rangle)^{1/3}$, with $\langle\rho\rangle$ the average dark matter density, and calculates the mass fraction in halos of mass $>M$ as the mass contained in regions with initial overdensity larger than a critical value $\delta_{\rm crit}$, after smoothing the dark matter density field over a length scale $R_M$,   
\begin{equation}
    F_M(\delta>\delta_{\rm crit})=\int_{\delta_{\rm crit}}^\infty \frac{dP_{R_M}}{d\delta}(1+\delta) d\delta.
\end{equation} The number of halos with mass $M$ is then simply obtained by differentiating over $F_M$, as
\begin{equation}
    \frac{{\rm d}n}{{\rm d}M} = \frac{\langle\rho\rangle}{M} \frac{\partial F_M(>M)}{\partial M}. 
\end{equation}
While the density contrast probability distribution $dP_{R_M}/d\delta$ can be easily calculated if the field is Gaussian, Refs.\cite{Vaquero:2018tib,Buschmann:2019icd} showed that direct simulations produced significant non-Gaussian tails in the Fourier distribution, see, e.g., Fig. 22 of \cite{Vaquero:2018tib}. Moreover, the density contrast fields can hardly be Gaussian, since by definition $\delta > -1$ and, its variance being of $\mathcal{O}(1)$, cannot be completely symmetric between overdensities and underdensities~\cite{Buschmann:2019icd}. However, we have checked explicitly that the distribution has the approximate shape of a \emph{log-normal} distribution with shifted mean,
\begin{equation}
    \frac{{\rm d}P_{R_M}}{{\rm d}\delta} = \frac{1}{\sqrt{2\pi}s_M(1+\delta)}\exp\left[{-\frac{(\log(1+\delta)+s_M^2/2)^2}{2 s_M^2}}\right], 
\end{equation} 
where $s_M^2=\log(1+\sigma_M^2)$ and $\sigma_M^2$ is the variance of the contrast fluctuations smoothed over distances $R_M$. The match improves with $R_M$, as smoothing erases correlations and reduces the variance. At large $R_M$, the distribution seems to approach a Gaussian. We can then approximately compute the HMF with the the log-normal distribution, obtaining
\begin{equation}
\label{HMFanal}
    \frac{{\rm d}n}{{\rm d}M}= \frac{\langle\rho\rangle}{M} \frac{{\rm d} s_M}{{\rm d} M} \frac{y_c+s_M^2/2}{\sqrt{2\pi}s_M^2}\exp\left[{-\frac{(y_c + s_M^2/2)^2}{2 s_M^2}}\right],~~~~y_c = \log(1+\delta_c),
\end{equation} where
$\delta_c \sim  (1+x^{-1})/(1+3x/2)$ (with $x=(1+z_{\rm eq})/(1+z)$) is the critical dark matter overdensity that will be collapsing at a given redshift,\footnote{An initially frozen dark matter overdensity $\delta_0=\rho_{\rm DM,0}/\langle\rho_{\rm DM}\rangle-1$ grows as $\delta_0(1+3 x/2)$ and will collapse once the dark matter density is $\mathcal{O}(1)$ larger than the average total energy density of the universe, i.e., radiation + matter. Since $\langle\rho_T\rangle =\langle\rho_{\rm DM}\rangle+\langle\rho_{\rm rad}\rangle=(1+x^{-1})\langle\rho_{\rm DM}\rangle$,  the critical value of $\delta_0$ must be $(1+x^{-1})/(1+3/2 x)$.} modified to include the radiation-domination era~\cite{Kolb:1994fi,Ellis:2020gtq}. In the white noise regime $\sigma_M^2$ should decrease as $R_M^{-3}\sim M^{-1}$, where fluctuations encompass many coherence lengths $R_M\gg L_1$ (equivalently $M\gg M_1\sim 10^{-13}~M_{\odot}$) and slowly level off to a constant value at $M\ll M_1$. 

\begin{figure}
    \centering
    \includegraphics[width=\textwidth]{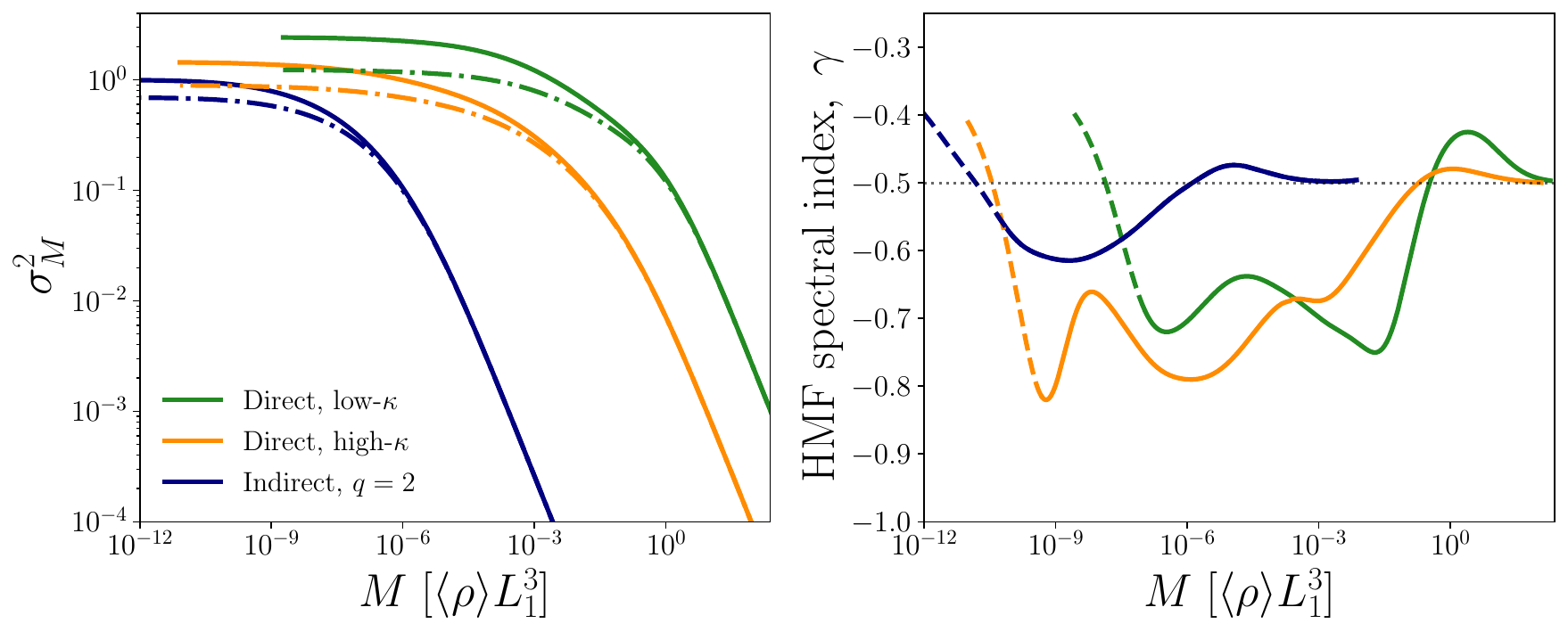}
    \caption{\emph{Left}: Variance of density contrast fluctuations as a function of the mass $M=\langle \rho\rangle (\sqrt{2\pi}R_M)^3$ after smoothing the power-spectrum with a Gaussian filter of width $\sigma=R_M$. The dot-dashed line shows $s_M^2=\log(1+\sigma_M^2)$. \emph{Right}: HMF spectral index $\gamma$ from Eq. \eqref{HMFanal}, neglecting the exponential cut-off.}
    \label{fig:slopeHMF}
\end{figure}

The normalised HMF ${\rm d}n/{\rm d}\log M$ is thus expected to be a smoothly varying function of $M$, namely  $({\rm d}s_M/{\rm d}M)s_{M}^{-2}$ multiplied by an exponential cut-off (since in practice $y_c>s_M^2$). Note that, as long as the deviations from Gaussianity are reasonably small, we can estimate $\sigma_M^2$ from the power spectrum as  
\begin{equation}
    \label{sigmaM2}
    \sigma_M^2 \simeq \int \frac{dk}{k}\Delta^2 e^{-k^2 R_M^2/2}. 
\end{equation}
Let us model the spectrum as a double power law $\Delta^2\sim \tilde{k}^3/(1+\tilde{k}^{3+\beta})$, with $\tilde{k}=kL_1$. 
In the white noise regime ($\tilde{k}<1$), it follows $\sigma_M^2\propto M^{-1}$ and therefore ${\rm d}n/{\rm d}\log M\propto M^{-1/2}$. Coincidentally, this is the same as the Press-Schechter result.
In the non-linear regime ($\tilde{k}>1$), on the other hand, $\sigma_M^2\sim M^0$ and therefore
${\rm d}n/{\rm d}\log M\propto 1/M^{1-\beta/3}$ for $0<\beta<2$ or ${\rm d}n/{\rm d}\log M\propto M^{-1/3}$ for $\beta>2$.
As seen in Fig.~\ref{fig:spectra}, estimating $\beta$ is not straightforward. However we expect $\beta=1$ for the indirect method,\footnote{This is because the energy spectrum from $q>1$ is $\partial\rho/\partial k\sim k^{-1}$ and we choose uncorrelated modes when we initialise the axion field.}
while larger values are expected for the direct methods, since they  reach values of $q\simeq 1$ during the simulations and have non-trivial correlations at small scales. 

We have computed the HMF spectral index (tilt) from the approximate equations \eqref{HMFanal} and \eqref{sigmaM2} and we show our results for $\sigma_M^2$ and $\gamma$ in Fig.~\ref{fig:slopeHMF} left and right, respectively. The value of $\gamma$ tends to $-0.5$ at large $M$, then decreases to values between $\gamma=-0.5$ and $\gamma=-1$ at intermediate masses and then increases again close to the artificial numerical cut-off (in dashed lines). Although all three simulations are qualitatively similar, there are important differences. Indeed, direct simulations predict larger fluctuations and have a much larger mass cut-off than the indirect ones. 
In the direct simulations, the intermediate slope is around $\gamma\sim -0.7$ for low-$\kappa$ and $\gamma\sim -0.75$ for high-$\kappa$, while for the indirect simulations we find $\gamma\sim -0.6$, close to the predicted $\gamma=-0.67$. As we will see, this semi-analytic modelling of $\gamma$ is confirmed by our numerical results from the N-body simulations. 

In addition, since the HMF is proportional to $(ds_M/dM)/s_M^2$, we expect it to be larger for the indirect simulation case at low masses and below the cut-off. However, the cut-off will happen at much lower masses and thus large-mass halos will be comparatively less frequent with respect to direct simulations. This is not surprising, given that in the indirect simulation large non-linearities at the QCD phase transition destroy the $\mathcal{O}(L_1)$ correlations in the spectrum, shifting the white-noise regime to smaller scales.  

\subsubsection{Numerical results}

\begin{figure}
    \centering
    \includegraphics[width=0.95\textwidth]{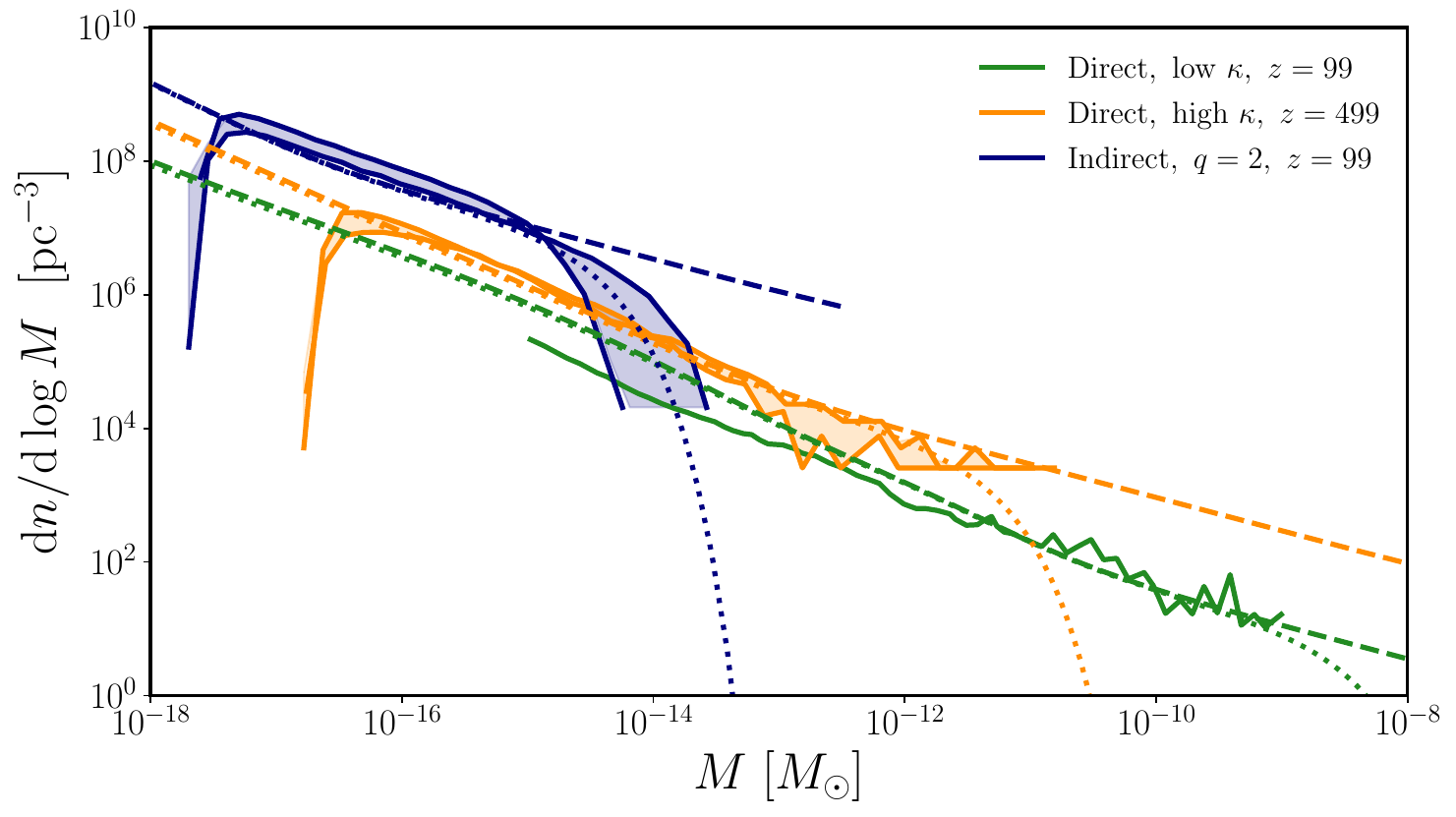}
    \caption{Minicluster halo mass function ${\rm d}n/{\rm d}\log M $ for the different methods and at the final simulation time. Direct simulations at low-$\kappa$ are taken from Ref.~\cite{Eggemeier:2019khm}. Dotted (dashed) lines correspond to our analytical prediction, Eq.~\eqref{HMFanal}, with (without) the exponential cutoff. }
    \label{fig:HMF}
\end{figure}

The minicluster HMF can be calculated directly either with the friend-of-friends algorithm or with the subfind halo finder.  In Fig.~\ref{fig:HMF} we show the resulting HMFs from direct and indirect simulations at the lowest redshift we can simulate accurately. We notice how direct simulations show slopes very close to $\gamma=-1$ at small masses and a slight increase over the low-mass trend at higher masses, which could be interpreted as a transition to the white-noise regime $\propto M^{-1/2}$. Since we  could not simulate values of $L/L_1>24$ for low$-\kappa$ and $L/L_1>3.5$ for high-$\kappa$ in the early stages, our simulations have only a few correlation lengths and cannot enter deeply into the white-noise regime with sufficient statistics (see, however Ref.~\cite{Xiao:2021nkb} for the study of the white-noise clustering). The slopes of the non-linear regime agree with the analytical expectations given above, but for a systematic factor of order of $\sim -0.1$ in all three cases, which is certainly acceptable given the crudeness (but simplicity) of the analytical estimate. Indeed, even the semi-analytical predictions of the HMF of \eqref{HMFanal} agree very well with the simulations (within a factor of $\sim 3$ or better).

It is also clear how the high-$\kappa$ simulations show more miniclusters in the low mass range with respect to low-$\kappa$ results\footnote{This still holds taking into account the redshift difference, see for example Fig. 2 of Ref.~\cite{Eggemeier:2019khm}.} and the transition to white-noise clearly happens at smaller masses in the high-$\kappa$ case, since the power-spectrum peaks at larger $k$ (see Fig.~\ref{fig:spectra}) and the correlation length is smaller. 
The indirect simulations show much more power at small scales, since the late nonlinearities push the non-linear scale to much higher $k$ than the characteristic scale $k_1$.
As a consequence, $\Delta^2\sim \mathcal{O}(1)$ at much smaller scales and the white noise is heavily reduced. As depicted in Fig.~\ref{fig:HMF}, the implication for miniclusters is that the exponential cut-off happens at much smaller masses. The slope of the low-mass regime is seen to be around $\gamma\simeq -0.68$, as expected analytically. 

\subsection{Density profiles}

\begin{figure}
    \centering
    \includegraphics[width=0.495\textwidth]{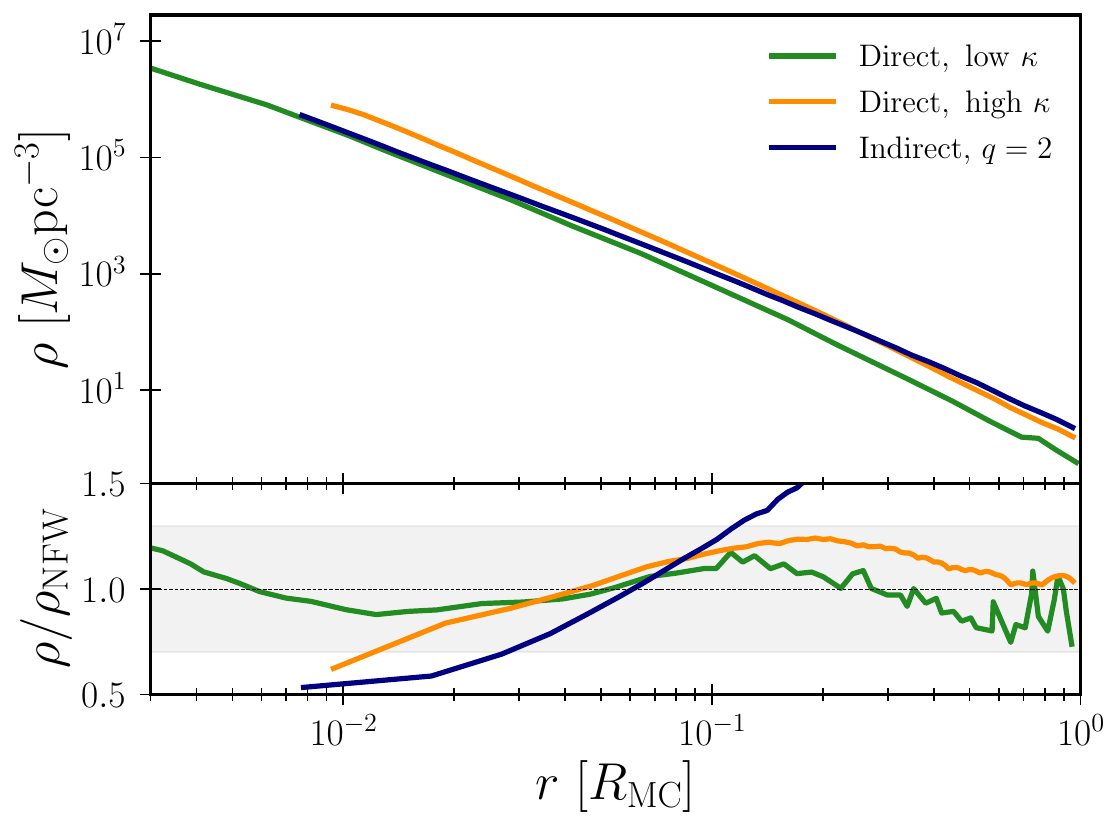}
    \includegraphics[width=0.495\textwidth]{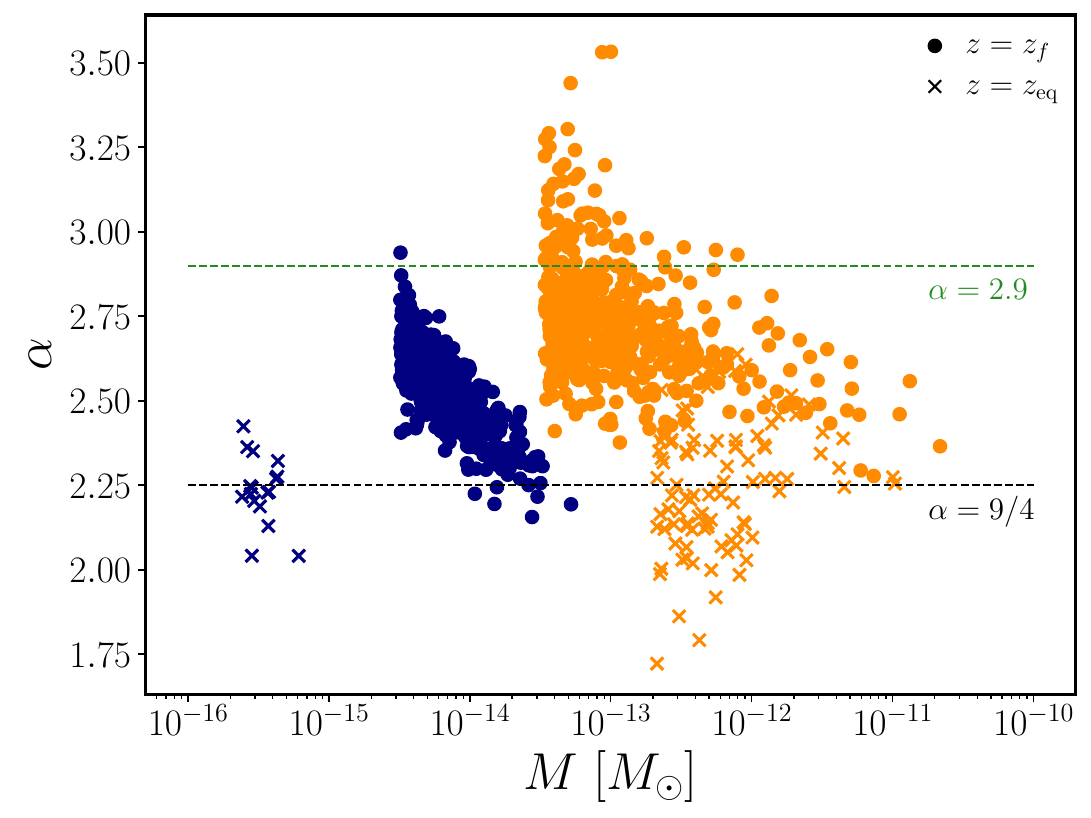}
    \caption{\emph{Left}: Averaged internal density profiles of the heaviest 500 miniclusters, normalised by their Virial radius. Results of direct simulations at low $\kappa$ are taken from Ref.~\cite{Eggemeier:2019khm}. On the bottom panel we show the deviation from the NFW fit. \emph{Right}: Distribution of power-law index fit, $\alpha$, for the heaviest miniclusters at $z_f$ and $z_{\rm eq}$. The medians found are $\alpha=2.71$ for direct high $\kappa$, and $\alpha=2.56$ for indirect simulations with $q=2$. We highlight the result from the analysis at low $\kappa$, $\alpha=2.9$~\cite{Ellis:2022grh}, and the expectation from self-similar infall, $\alpha=9/4$~\cite{Bertschinger:1985pd}.}
    \label{fig:profs}
\end{figure}

In the case of direct simulations at low $\kappa$, radial density profiles of miniclusters and their halos have been analysed in the N-body simulation of Ref.~\cite{Eggemeier:2019khm} and using the Peak-Patch method~\cite{Ellis:2020gtq}. A recent analysis on the same simulation has been also performed in Ref.~\cite{Ellis:2022grh}, which considered the NFW~\cite{Navarro:1996gj} and single power-law (PL) models to fit the data. The main takeaway from that analysis is that, due to limited spatial resolution to reach the best-fit NFW scale radius, 80\% of miniclusters can be also fitted with a PL behaviour $\rho\sim r^{-\alpha}$, with $\alpha=2.9$ the average PL index. This means that for these miniclusters the density profile estimation is essentially inconclusive. We estimated the density profiles from our new suite of simulations by analysing the largest 500 miniclusters in our numerical realisations and found considerable differences between direct and indirect simulations.
As seen in the left panel of Fig.~\ref{fig:profs}, direct simulations are in good agreement with the NFW modelling, although we do not resolve the scale radius for all the miniclusters analysed.
Given the already discussed differences in the initial spectrum $\Delta^2$, the only difference between low-$\kappa$ and high-$\kappa$ direct simulations is that the latter has miniclusters with typically larger densities.
Indirect simulations on the other hand do not reproduce the NFW profile and typically show power-law profiles with an average index $\alpha=2.56$, see the right panel of Fig.~\ref{fig:profs}.
This is not accidental, given that indirect simulations show less hierarchical merging with respect to direct simulations, and many miniclusters that form at the earliest times survive as isolated halos.
As pointed out in the analysis of  Ref.~\cite{Ellis:2022grh}, we decide to also fit direct simulations with a power law model and found an average index $\alpha=2.71$, larger than the indirect case, but smaller than the low-$\kappa$ results.
In the left panel of Fig.~\ref{fig:profs} we show the distribution of power-law best fits as a function of the minicluster mass, highlighting the variance within each method. We also provide a comparison of the density profiles between matter-radiation equality $z_{\rm eq}=3400$ and the final simulation time. Indeed, we observe a slight time dependence towards larger values of $\alpha$, and an agreement with the result $\alpha=9/4$ at $z_{\rm eq}$.

\section{Present-day implications}\label{sec:impl}

As already mentioned, our numerical analyses cannot describe the distribution of miniclusters at the present time. The subsequent hierarchical merging into larger minicluster halos should be combined with late-time disruption dynamics. Therefore, a reasonable extrapolation of our results is beyond the scope of this work.
Nevertheless, we briefly comment on the implications for direct and indirect detection for the minicluster scenario.
Recent Monte Carlo simulations~\cite{Kavanagh:2020gcy} have shown how miniclusters with PL profiles are generally more resilient to tidal disruption than miniclusters with NFW profiles. This implies that the survival probability of miniclusters strongly depends on the internal structure, see also Ref.~\cite{Dandoy:2022prp}.
In addition, recent work \cite{Shen:2022ltx}, while mostly focusing on NFW modelling, emphasises the importance of studying the system with a realistic minicluster concentration distribution. In our work we have not attempted to reach a better resolution to determine the internal density profiles. However we have shown how the different methods to simulate early Universe dynamics lead to generally different profiles and mass ranges.
This in turn has some clear implications for indirect axion dark matter probes. 
Indeed, for low-$\kappa$ axions to provide miniclusters which can be detected via microlensing requires a small axion mass window around $m_a\sim {\rm meV}$, in disagreement with most dark matter mass predictions from direct simulations, both low-$\kappa$ and high-$\kappa$.
In our new simulations, both direct high-$\kappa$ and indirect, $q=2$ simulations find miniclusters with a much smaller mass distribution, which is therefore less likely to lead to microlensing events.
In particular, indirect simulations show a maximum minicluster mass of $M\sim 3\times 10^{-14}~M_{\odot}$ at $z=99$, arguably too far from the minimum mass $M_{\rm min}\simeq 10^{-11}~M_{\odot}$ required by wave optics~\cite{Ellis:2022grh}, even for results at $z=0$. 

On the direct detection side, if axions constitute the dark matter and the PQ symmetry breaks after the end of inflation, axion haloscopes need to take into account the dark matter substructure on very small scales. The expected signal, being proportional to the local value of the energy density $\rho_a({\bf x})$, would indeed have considerable spatial gradients on scales probed by experiments $\ell\lesssim {\rm mpc}$.
For the direct low-$\kappa$ case, Ref.~\cite{Eggemeier:2022hqa} first estimated the typical correction one has to apply to the experimental sensitivity, which assumes the inferred average value $\rho_{\rm DM}=0.4~{\rm GeV}~{\rm cm}^{-3}$,%
\footnote{This value is typically used in the experimental analysis and to constraint the $g-m_a$ parameter space. However the estimate varies in the literature~\cite{Read:2014qva}.}
in the case where haloscopes turn out to be probing an underdense region.
The loss in the $g\propto\rho_a^{-1/2}$ sensitivity, with $g$ the given axion coupling used by the experiment (see a collection of the data ~\cite{AxionLimits}), was estimated to be a factor of $C\sim 3.5$~\cite{Eggemeier:2022hqa}. This correction can be taken as a conservative estimate, since the subsequent dynamics to $z=0$ can only lead to an energy enhancement of underdense regions because of tidal minicluster streams. In our new simulations, see Fig.~\ref{fig:budget} (right), we have shown that the typical void density is comparable for the different methods at the final simulation time. In particular, indirect simulations show marginally denser voids, at least on average, and at $z=99$ the void energy approaches a constant value, as such regions cannot further comovingly expand in space. The results at high-$\kappa$ are unfortunately limited to $z=499$, but show a quite good agreement with results at low-$\kappa$. This implies that the \emph{conservative} worst-case-scenario estimate of Ref.~\cite{Eggemeier:2022hqa} generally holds independently of the initial conditions.
On the other hand, given the differences in the internal structure of miniclusters among the methods simulated, we expect the late time tidal disruption dynamics to generally induce different results in the small-scale distribution of dark matter.

\section{Conclusions}\label{sec:conclusions}

In this work we have investigated how the numerical modelling of early Universe dynamics affects the phenomenology of axion miniclusters and the ensuing implications for experimental searches. We have performed a suite of simulations where we evolved a realistic field configuration from the early Universe epoch, $z\sim 10^{13}$, until a final redshift $z\ll z_{\rm eq}$ well after matter-radiation equality, using a mixture of numerical methods and different codes.
We have compared different approaches to overcome the technical obstacles involved in simulating strings with physical value of the tension, and therefore improved upon previous simulations of axion miniclusters.

We find that direct simulations with high tension strings from Ref.~\cite{Klaer:2017ond,Klaer:2017qhr} show large correlations at larger $k$ values with respect to the case at small tension, leading to a population of miniclusters with smaller masses and larger densities. We also find good agreement in the HMF spectral index $\gamma$, both from our analytical expectations and the numerical results. 

Indirect simulations with $q>1$ also show density fluctuations at much smaller spatial scales, leading to even lower-mass miniclusters. The result at late times shows a distribution of more isolated miniclusters (i.e., with fewer hierarchical mergers)  compared to direct simulations. We find differences in the HMF spectral index and the internal density profiles, the latter being closer to PL rather than NFW.

Nevertheless, despite the differences in the HMF and density profiles, the void densities are very similar between different simulation types, which affirms the solidity of the estimate in Ref. ~\cite{Eggemeier:2022hqa}. As discussed, the present-day spatial distribution of miniclusters is beyond the scope of this work and will ultimately need to be carefully taken into consideration through dedicated numerical studies. 

\section*{Acknowledgements}

We thank Mathieu Kaltschmidt, Doddy Marsh, Ciaran O'Hare and Yvonne Wong for feedback on this draft and useful conversations. G.P. and J.R. would like to acknowledge GGI for the hospitality during early stages of this work. This research was undertaken using the HPC systems \emph{Gadi} from the National Computational Infrastructure (NCI) supported by the Australian Government, and \emph{Raven} at the Max Planck Computing \& data facility (MPCDF). We acknowledge the use of the \texttt{Pylians}\footnote{\href{https://github.com/franciscovillaescusa/Pylians3}{https://github.com/franciscovillaescusa/Pylians3}} library for the analysis. This article is based upon work from COST Action COSMIC WISPers CA21106, 
supported by COST (European Cooperation in Science and Technology). G.P.\ is supported by an University International Postgraduate Award from UNSW. J.R.\ is supported by Grant PGC2022-126078NB-C21 funded by MCIN/AEI/ 10.13039/501100011033 and “ERDF A way of making Europe”, and J.R.\ and A.V.\ are supporterd by Grant  DGA-FSE grant 2020-E21-17R Aragon Government and the European
Union - NextGenerationEU Recovery and Resilience Program on `Astrofísica y Física de Altas Energías' CEFCA-CAPA-ITAINNOVA. A.V.\ is further supported by AEI (Spain) under Grant No. RYC2020-030244-I / AEI / 10.13039/501100011033. The work of K.S.\ is supported by Leading Initiative for Excellent Young Researchers (LEADER), the Ministry of Education, Culture, Sports, Science, and Technology (MEXT), Japan. G.M.\ is supported by the Technische Universit\"at Darmstadt through its Institut f\"ur Kernphysik.

\appendix 

\bibliography{axions.bib}

\providecommand{\href}[2]{#2}\begingroup\raggedright\begin{thebibliography}{10}

\bibitem{Vaquero:2018tib}
A.~Vaquero, J.~Redondo and J.~Stadler, \emph{{Early seeds of axion
  miniclusters}},
  \href{https://doi.org/10.1088/1475-7516/2019/04/012}{\emph{JCAP} {\bfseries
  04} (2019) 012} [\href{https://arxiv.org/abs/1809.09241}{{\ttfamily
  1809.09241}}].

\bibitem{Klaer:2017ond}
V.~B. Klaer and G.~D. Moore, \emph{{The dark-matter axion mass}},
  \href{https://doi.org/10.1088/1475-7516/2017/11/049}{\emph{JCAP} {\bfseries
  1711} (2017) 049} [\href{https://arxiv.org/abs/1708.07521}{{\ttfamily
  1708.07521}}].

\bibitem{Gorghetto:2020qws}
M.~Gorghetto, E.~Hardy and G.~Villadoro, \emph{{More Axions from Strings}},
  \href{https://doi.org/10.21468/SciPostPhys.10.2.050}{\emph{SciPost Phys.}
  {\bfseries 10} (2021) 050}
  [\href{https://arxiv.org/abs/2007.04990}{{\ttfamily 2007.04990}}].

\bibitem{Peccei:1977hh}
R.~D. Peccei and H.~R. Quinn, \emph{{CP conservation in the presence of
  instantons}}, \href{https://doi.org/10.1103/PhysRevLett.38.1440}{\emph{Phys.
  Rev. Lett.} {\bfseries 38} (1977) 1440}. [,328(1977)].

\bibitem{Bertone:2016nfn}
G.~Bertone and D.~Hooper, \emph{{History of dark matter}},
  \href{https://doi.org/10.1103/RevModPhys.90.045002}{\emph{Rev. Mod. Phys.}
  {\bfseries 90} (2018) 045002}
  [\href{https://arxiv.org/abs/1605.04909}{{\ttfamily 1605.04909}}].

\bibitem{Preskill:1982cy}
J.~Preskill, M.~B. Wise and F.~Wilczek, \emph{{Cosmology of the Invisible
  Axion}}, \href{https://doi.org/10.1016/0370-2693(83)90637-8}{\emph{Phys.
  Lett. B} {\bfseries 120} (1983) 127}.

\bibitem{Abbott:1982af}
L.~F. Abbott and P.~Sikivie, \emph{{A Cosmological Bound on the Invisible
  Axion}}, \href{https://doi.org/10.1016/0370-2693(83)90638-X}{\emph{Phys.
  Lett. B} {\bfseries 120} (1983) 133}.

\bibitem{Dine:1981rt}
M.~Dine, W.~Fischler and M.~Srednicki, \emph{{A simple solution to the Strong
  CP Problem with a harmless axion}},
  \href{https://doi.org/10.1016/0370-2693(81)90590-6}{\emph{Phys. Lett. B}
  {\bfseries 104} (1981) 199}.

\bibitem{Dine:1982ah}
M.~Dine and W.~Fischler, \emph{{The Not So Harmless Axion}},
  \href{https://doi.org/10.1016/0370-2693(83)90639-1}{\emph{Phys. Lett. B}
  {\bfseries 120} (1983) 137}.

\bibitem{Turner:1983he}
M.~S. Turner, \emph{{Coherent Scalar Field Oscillations in an Expanding
  Universe}}, \href{https://doi.org/10.1103/PhysRevD.28.1243}{\emph{Phys. Rev.
  D} {\bfseries 28} (1983) 1243}.

\bibitem{Turner:1985si}
M.~S. Turner, \emph{{Cosmic and Local Mass Density of Invisible Axions}},
  \href{https://doi.org/10.1103/PhysRevD.33.889}{\emph{Phys. Rev. D} {\bfseries
  33} (1986) 889}.

\bibitem{DiLuzio:2020wdo}
L.~Di~Luzio, M.~Giannotti, E.~Nardi and L.~Visinelli, \emph{{The landscape of
  QCD axion models}},
  \href{https://doi.org/10.1016/j.physrep.2020.06.002}{\emph{Phys. Rept.}
  {\bfseries 870} (2020) 1} [\href{https://arxiv.org/abs/2003.01100}{{\ttfamily
  2003.01100}}].

\bibitem{Irastorza:2018dyq}
I.~G. Irastorza and J.~Redondo, \emph{{New experimental approaches in the
  search for axion-like particles}},
  \href{https://doi.org/10.1016/j.ppnp.2018.05.003}{\emph{Prog. Part. Nucl.
  Phys.} {\bfseries 102} (2018) 89}
  [\href{https://arxiv.org/abs/1801.08127}{{\ttfamily 1801.08127}}].

\bibitem{Hogan:1988mp}
C.~J. Hogan and M.~J. Rees, \emph{{Axion miniclusters}},
  \href{https://doi.org/10.1016/0370-2693(88)91655-3}{\emph{Phys. Lett. B}
  {\bfseries 205} (1988) 228}.

\bibitem{Kolb:1994fi}
E.~W. Kolb and I.~I. Tkachev, \emph{{Large amplitude isothermal fluctuations
  and high density dark matter clumps}},
  \href{https://doi.org/10.1103/PhysRevD.50.769}{\emph{Phys. Rev. D} {\bfseries
  50} (1994) 769} [\href{https://arxiv.org/abs/astro-ph/9403011}{{\ttfamily
  astro-ph/9403011}}].

\bibitem{Kibble:1976sj}
T.~W.~B. Kibble, \emph{{Topology of Cosmic Domains and Strings}},
  \href{https://doi.org/10.1088/0305-4470/9/8/029}{\emph{J. Phys. A} {\bfseries
  9} (1976) 1387}.

\bibitem{Gorghetto:2018myk}
M.~Gorghetto, E.~Hardy and G.~Villadoro, \emph{{Axions from Strings: the
  Attractive Solution}},
  \href{https://doi.org/10.1007/JHEP07(2018)151}{\emph{JHEP} {\bfseries 07}
  (2018) 151} [\href{https://arxiv.org/abs/1806.04677}{{\ttfamily
  1806.04677}}].

\bibitem{Hiramatsu:2012gg}
T.~Hiramatsu, M.~Kawasaki, K.~Saikawa and T.~Sekiguchi, \emph{{Production of
  dark matter axions from collapse of string-wall systems}},
  \href{https://doi.org/10.1103/PhysRevD.85.105020}{\emph{Phys. Rev. D}
  {\bfseries 85} (2012) 105020}
  [\href{https://arxiv.org/abs/1202.5851}{{\ttfamily 1202.5851}}]. [Erratum:
  Phys.Rev.D 86, 089902 (2012)].

\bibitem{Fleury:2015aca}
L.~Fleury and G.~D. Moore, \emph{{Axion dark matter: strings and their cores}},
  \href{https://doi.org/10.1088/1475-7516/2016/01/004}{\emph{JCAP} {\bfseries
  01} (2016) 004} [\href{https://arxiv.org/abs/1509.00026}{{\ttfamily
  1509.00026}}].

\bibitem{Klaer:2017qhr}
V.~B. Klaer and G.~D. Moore, \emph{{How to simulate global cosmic strings with
  large string tension}},
  \href{https://doi.org/10.1088/1475-7516/2017/10/043}{\emph{JCAP} {\bfseries
  10} (2017) 043} [\href{https://arxiv.org/abs/1707.05566}{{\ttfamily
  1707.05566}}].

\bibitem{Buschmann:2019icd}
M.~Buschmann, J.~W. Foster and B.~R. Safdi, \emph{{Early-Universe Simulations
  of the Cosmological Axion}},
  \href{https://doi.org/10.1103/PhysRevLett.124.161103}{\emph{Phys. Rev. Lett.}
  {\bfseries 124} (2020) 161103}
  [\href{https://arxiv.org/abs/1906.00967}{{\ttfamily 1906.00967}}].

\bibitem{Buschmann:2021sdq}
M.~Buschmann, J.~W. Foster, A.~Hook, A.~Peterson, D.~E. Willcox, W.~Zhang and
  B.~R. Safdi, \emph{{Dark Matter from Axion Strings with Adaptive Mesh
  Refinement}},  \href{https://arxiv.org/abs/2108.05368}{{\ttfamily
  2108.05368}}.

\bibitem{Harari:1987ht}
D.~Harari and P.~Sikivie, \emph{{On the Evolution of Global Strings in the
  Early Universe}},
  \href{https://doi.org/10.1016/0370-2693(87)90032-3}{\emph{Phys. Lett. B}
  {\bfseries 195} (1987) 361}.

\bibitem{Hagmann:1990tj}
C.~Hagmann, P.~Sikivie, N.~S. Sullivan and D.~B. Tanner, \emph{{Results from a
  search for cosmic axions}},
  \href{https://doi.org/10.1103/PhysRevD.42.1297}{\emph{Phys. Rev. D}
  {\bfseries 42} (1990) 1297}.

\bibitem{Dine:2020pds}
M.~Dine, N.~Fernandez, A.~Ghalsasi and H.~H. Patel, \emph{{Comments on axions,
  domain walls, and cosmic strings}},
  \href{https://doi.org/10.1088/1475-7516/2021/11/041}{\emph{JCAP} {\bfseries
  11} (2021) 041} [\href{https://arxiv.org/abs/2012.13065}{{\ttfamily
  2012.13065}}].

\bibitem{Kawasaki:2014sqa}
M.~Kawasaki, K.~Saikawa and T.~Sekiguchi, \emph{{Axion dark matter from
  topological defects}},
  \href{https://doi.org/10.1103/PhysRevD.91.065014}{\emph{Phys. Rev. D}
  {\bfseries 91} (2015) 065014}
  [\href{https://arxiv.org/abs/1412.0789}{{\ttfamily 1412.0789}}].

\bibitem{OHare:2021zrq}
C.~A.~J. O'Hare, G.~Pierobon, J.~Redondo and Y.~Y.~Y. Wong, \emph{{Simulations
  of axionlike particles in the postinflationary scenario}},
  \href{https://doi.org/10.1103/PhysRevD.105.055025}{\emph{Phys. Rev. D}
  {\bfseries 105} (2022) 055025}
  [\href{https://arxiv.org/abs/2112.05117}{{\ttfamily 2112.05117}}].

\bibitem{Davis:1985pt}
R.~L. Davis, \emph{{Goldstone Bosons in String Models of Galaxy Formation}},
  \href{https://doi.org/10.1103/PhysRevD.32.3172}{\emph{Phys. Rev. D}
  {\bfseries 32} (1985) 3172}.

\bibitem{Davis:1986xc}
R.~L. Davis, \emph{{Cosmic Axions from Cosmic Strings}},
  \href{https://doi.org/10.1016/0370-2693(86)90300-X}{\emph{Phys. Lett. B}
  {\bfseries 180} (1986) 225}.

\bibitem{Battye:1993jv}
R.~A. Battye and E.~P.~S. Shellard, \emph{{Global string radiation}},
  \href{https://doi.org/10.1016/0550-3213(94)90573-8}{\emph{Nucl. Phys. B}
  {\bfseries 423} (1994) 260}
  [\href{https://arxiv.org/abs/astro-ph/9311017}{{\ttfamily
  astro-ph/9311017}}].

\bibitem{Battye:1994au}
R.~A. Battye and E.~P.~S. Shellard, \emph{{Axion string constraints}},
  \href{https://doi.org/10.1103/PhysRevLett.73.2954}{\emph{Phys. Rev. Lett.}
  {\bfseries 73} (1994) 2954}
  [\href{https://arxiv.org/abs/astro-ph/9403018}{{\ttfamily
  astro-ph/9403018}}]. [Erratum: Phys.Rev.Lett. 76, 2203--2204 (1996)].

\bibitem{Eggemeier:2019khm}
B.~Eggemeier, J.~Redondo, K.~Dolag, J.~C. Niemeyer and A.~Vaquero, \emph{{First
  Simulations of Axion Minicluster Halos}},
  \href{https://doi.org/10.1103/PhysRevLett.125.041301}{\emph{Phys. Rev. Lett.}
  {\bfseries 125} (2020) 041301}
  [\href{https://arxiv.org/abs/1911.09417}{{\ttfamily 1911.09417}}].

\bibitem{Springel:2020plp}
V.~Springel, R.~Pakmor, O.~Zier and M.~Reinecke, \emph{{Simulating cosmic
  structure formation with the gadget-4 code}},
  \href{https://doi.org/10.1093/mnras/stab1855}{\emph{Mon. Not. Roy. Astron.
  Soc.} {\bfseries 506} (2021) 2871}
  [\href{https://arxiv.org/abs/2010.03567}{{\ttfamily 2010.03567}}].

\bibitem{Eggemeier:2022hqa}
B.~Eggemeier, C.~A.~J. O'Hare, G.~Pierobon, J.~Redondo and Y.~Y.~Y. Wong,
  \emph{{Axion minivoids and implications for direct detection}},
  \href{https://doi.org/10.1103/PhysRevD.107.083510}{\emph{Phys. Rev. D}
  {\bfseries 107} (2023) 083510}
  [\href{https://arxiv.org/abs/2212.00560}{{\ttfamily 2212.00560}}].

\bibitem{Kolb:1993hw}
E.~W. Kolb and I.~I. Tkachev, \emph{{Nonlinear axion dynamics and formation of
  cosmological pseudosolitons}},
  \href{https://doi.org/10.1103/PhysRevD.49.5040}{\emph{Phys. Rev. D}
  {\bfseries 49} (1994) 5040}
  [\href{https://arxiv.org/abs/astro-ph/9311037}{{\ttfamily
  astro-ph/9311037}}].

\bibitem{Dokuchaev}
V.~I. {Dokuchaev}, Y.~N. {Eroshenko} and I.~I. {Tkachev}, \emph{{Destruction of
  axion miniclusters in the Galaxy}},
  \href{https://doi.org/10.1134/S1063776117080039}{\emph{Soviet Journal of
  Experimental and Theoretical Physics} {\bfseries 125} (2017) 434}
  [\href{https://arxiv.org/abs/1710.09586}{{\ttfamily 1710.09586}}].

\bibitem{Tinyakov:2015cgg}
P.~Tinyakov, I.~Tkachev and K.~Zioutas, \emph{{Tidal streams from axion
  miniclusters and direct axion searches}},
  \href{https://doi.org/10.1088/1475-7516/2016/01/035}{\emph{JCAP} {\bfseries
  01} (2016) 035} [\href{https://arxiv.org/abs/1512.02884}{{\ttfamily
  1512.02884}}].

\bibitem{Kavanagh:2020gcy}
B.~J. Kavanagh, T.~D.~P. Edwards, L.~Visinelli and C.~Weniger, \emph{{Stellar
  Disruption of Axion Miniclusters in the Milky Way}},
  \href{https://arxiv.org/abs/2011.05377}{{\ttfamily 2011.05377}}.

\bibitem{Dandoy:2022prp}
V.~Dandoy, T.~Schwetz and E.~Todarello, \emph{{A self-consistent wave
  description of axion miniclusters and their survival in the galaxy}},
  \href{https://doi.org/10.1088/1475-7516/2022/09/081}{\emph{JCAP} {\bfseries
  09} (2022) 081} [\href{https://arxiv.org/abs/2206.04619}{{\ttfamily
  2206.04619}}].

\bibitem{Shen:2022ltx}
X.~Shen, H.~Xiao, P.~F. Hopkins and K.~M. Zurek, \emph{{Disruption of Dark
  Matter Minihaloes in the Milky Way environment: Implications for Axion
  Miniclusters and Early Matter Domination}},
  \href{https://arxiv.org/abs/2207.11276}{{\ttfamily 2207.11276}}.

\bibitem{Borsanyi:2016ksw}
S.~Borsanyi et~al., \emph{{Calculation of the axion mass based on
  high-temperature lattice quantum chromodynamics}},
  \href{https://doi.org/10.1038/nature20115}{\emph{Nature} {\bfseries 539}
  (2016) 69} [\href{https://arxiv.org/abs/1606.07494}{{\ttfamily 1606.07494}}].

\bibitem{Zeldovich:1974uw}
Y.~B. Zeldovich, I.~Y. Kobzarev and L.~B. Okun, \emph{{Cosmological
  Consequences of the Spontaneous Breakdown of Discrete Symmetry}}, {\emph{Zh.
  Eksp. Teor. Fiz.} {\bfseries 67} (1974) 3}.

\bibitem{Sikivie:1982qv}
P.~Sikivie, \emph{{Of Axions, Domain Walls and the Early Universe}},
  \href{https://doi.org/10.1103/PhysRevLett.48.1156}{\emph{Phys. Rev. Lett.}
  {\bfseries 48} (1982) 1156}.

\bibitem{Press:1972zz}
W.~H. Press and S.~A. Teukolsky, \emph{{Floating Orbits, Superradiant
  Scattering and the Black-hole Bomb}},
  \href{https://doi.org/10.1038/238211a0}{\emph{Nature} {\bfseries 238} (1972)
  211}.

\bibitem{Press:1973iz}
W.~H. Press and P.~Schechter, \emph{{Formation of galaxies and clusters of
  galaxies by selfsimilar gravitational condensation}},
  \href{https://doi.org/10.1086/152650}{\emph{Astrophys. J.} {\bfseries 187}
  (1974) 425}.

\bibitem{RS}
J.~Redondo, K.~Saikawa and A.~Vaquero, \emph{{In preparation}}, {\emph{{}}
  (2023) }.

\bibitem{Enander:2017ogx}
J.~Enander, A.~Pargner and T.~Schwetz, \emph{{Axion minicluster power spectrum
  and mass function}},
  \href{https://doi.org/10.1088/1475-7516/2017/12/038}{\emph{JCAP} {\bfseries
  12} (2017) 038} [\href{https://arxiv.org/abs/1708.04466}{{\ttfamily
  1708.04466}}].

\bibitem{Ellis:2020gtq}
D.~Ellis, D.~J.~E. Marsh and C.~Behrens, \emph{{Axion Miniclusters Made Easy}},
  \href{https://doi.org/10.1103/PhysRevD.103.083525}{\emph{Phys. Rev. D}
  {\bfseries 103} (2021) 083525}
  [\href{https://arxiv.org/abs/2006.08637}{{\ttfamily 2006.08637}}].

\bibitem{Eroncel:2022efc}
C.~Er\"oncel and G.~Servant, \emph{{ALP dark matter mini-clusters from kinetic
  fragmentation}},
  \href{https://doi.org/10.1088/1475-7516/2023/01/009}{\emph{JCAP} {\bfseries
  01} (2023) 009} [\href{https://arxiv.org/abs/2207.10111}{{\ttfamily
  2207.10111}}].

\bibitem{Chatrchyan:2023cmz}
A.~Chatrchyan, C.~Er\"oncel, M.~Koschnitzke and G.~Servant, \emph{{ALP dark
  matter with non-periodic potentials: parametric resonance, halo formation and
  gravitational signatures}},
  \href{https://arxiv.org/abs/2305.03756}{{\ttfamily 2305.03756}}.

\bibitem{Xiao:2021nkb}
H.~Xiao, I.~Williams and M.~McQuinn, \emph{{Simulations of axion minihalos}},
  \href{https://doi.org/10.1103/PhysRevD.104.023515}{\emph{Phys. Rev. D}
  {\bfseries 104} (2021) 023515}
  [\href{https://arxiv.org/abs/2101.04177}{{\ttfamily 2101.04177}}].

\bibitem{Ellis:2022grh}
D.~Ellis, D.~J.~E. Marsh, B.~Eggemeier, J.~Niemeyer, J.~Redondo and K.~Dolag,
  \emph{{Structure of axion miniclusters}},
  \href{https://doi.org/10.1103/PhysRevD.106.103514}{\emph{Phys. Rev. D}
  {\bfseries 106} (2022) 103514}
  [\href{https://arxiv.org/abs/2204.13187}{{\ttfamily 2204.13187}}].

\bibitem{Bertschinger:1985pd}
E.~Bertschinger, \emph{{Self - similar secondary infall and accretion in an
  Einstein-de Sitter universe}},
  \href{https://doi.org/10.1086/191028}{\emph{Astrophys. J. Suppl.} {\bfseries
  58} (1985) 39}.

\bibitem{Navarro:1996gj}
J.~F. Navarro, C.~S. Frenk and S.~D.~M. White, \emph{{A Universal density
  profile from hierarchical clustering}},
  \href{https://doi.org/10.1086/304888}{\emph{Astrophys. J.} {\bfseries 490}
  (1997) 493} [\href{https://arxiv.org/abs/astro-ph/9611107}{{\ttfamily
  astro-ph/9611107}}].

\bibitem{Read:2014qva}
J.~I. Read, \emph{{The Local Dark Matter Density}},
  \href{https://doi.org/10.1088/0954-3899/41/6/063101}{\emph{J. Phys. G}
  {\bfseries 41} (2014) 063101}
  [\href{https://arxiv.org/abs/1404.1938}{{\ttfamily 1404.1938}}].

\bibitem{AxionLimits}
C.~O'Hare, \emph{cajohare/axionlimits: Axionlimits},  July, 2020.
\newblock 10.5281/zenodo.3932430.

\end{thebibliography}\endgroup
\bibliographystyle{bibi}

\end{document}